\documentclass[%
 reprint,
 amsmath,amssymb,
 aps,
]{revtex4-2}

\usepackage{graphicx}% Include figure files
\usepackage{dcolumn}% Align table columns on decimal point
\usepackage{bm}% bold math
\usepackage{hyperref}% add hypertext capabilities
\begin{document}

\preprint{APS/123-QED}

\title{Towards the Double Copy Formulation for the Abelian Sector of Heterotic Double Field Theory}% Force line breaks with \\

\author{Rasim Yılmaz}%
\email{ryilmaz@metu.edu.tr}
\affiliation{%
Department of Physics, Faculty of Arts and Sciences\\
Middle East Technical University, 06800, Ankara, Türkiye
}%

\date{\today}% It is always \today, today,

\begin{abstract}
 It has been shown that double field theory (DFT) action can be constructed in a perturbative manner up to cubic order by using double copy of Yang-Mills. This construction can also be extended by starting with higher-derivative Yang-Mills theory to obtain higher-derivative DFT action. In this work, we extend this classical double copy formulation to the Abelian subsector of heterotic DFT, starting with a Yang-Mills theory that includes a charged scalar field, namely ($YM+\phi^3$) theory. The action we constructed by a double copy procedure exactly matches with the Abelian sector of heterotic DFT action up to cubic order.  

\end{abstract}

\maketitle

%\tableofcontents
\section{Introduction}
The Bern-Carrasco-Johansson (BCJ) double copy \cite{bcj,bcj2} is a remarkable correspondence between gauge theories, such as Yang-Mills theory, and gravity. Originally discovered in the study of scattering amplitudes, this duality reveals deep structural similarities between these seemingly different physical theories. The double copy states that gravitational amplitudes can be obtained by "squaring" gauge theory amplitudes, providing a powerful method for simplifying complex gravity calculations. \\

The origins of the double copy can be traced back to the Kawai-Lewellen-Tye (KLT) relations \cite{klt}, which connect string theory amplitudes for closed and open strings. In a field theory setting, this idea was refined into the BCJ duality, which expresses gauge theory amplitudes in a form that directly manifests their connection to gravity. Besides being in the center of modern amplitude program, the implications of double copy also extend to the classical solutions (see \cite{Bern2010,johan,johan2,Luna,Carrillo,abbas,adamo, expform, Leekerr, Cholee, Kim}), and even to the non-commutative theories \cite{lescanononcom}. \\

An interesting application of double copy idea on the action level is introduced in the works \cite{jaram}, \cite{dario}, which show that perturbative double field theory (DFT) action can be written as double copy of Yang-Mills theory. DFT is a framework that makes T-duality manifest by treating both the usual spacetime coordinates and their duals on an equal footing (see \cite{hulldft,backindepdft,hohmdft,aldaz}). This interesting relation between double copy and DFT is also studied in the level of classical solutions \cite{Lee,Cho,lescanohetero}, higher derivative Yang-Mills theories \cite{Lescano:2023pai, Lescano:2024gma, Lescano:2024lwn, rasim}, and gauge algebra structures \cite{Bonezzi, Bonezzi2, Bonezzi3, Bonezzi4}. \\

In another interesting work, the authors extend the construction of DFT action by the double copy procedure to the higher derivative Yang-Mills theories \cite{Lescano:2024gma}. Inspired by the work \cite{johan}, which states that conformal gravity amplitudes can be obtained by a double copy of pure Yang-Mills theory and dimension-six theory (Yang-Mills including a charged scalar), the authors of \cite{Lescano:2024gma} try to construct a conformal DFT action in a perturbative manner. Then, they introduce a double copy procedure which starts with the dimension-six theory introduced in \cite{johan}. \\

Following the same logic as in \cite{Lescano:2024gma}, in this work, we construct a double copy formulation for the Abelian subsector of heterotic DFT. Heterotic DFT is an extension of usual DFT formulation of the low-energy theory of the closed bosonic string to the heterotic string \cite{Hohmhetero, Hohmframe}. There are already some works on the literature about the double copy construction of heterotic DFT \cite{lescanohetero,Lescanohetero2,Chohetero}, which all use the Kerr-Schild ansatz. Kerr-Schild formalism is a powerful tool for solving the Einstein equation by reducing it to linear equations. The important consequence is that a class
of solutions to the vacuum Einstein equation can be represented by the square of solutions to Maxwell theory \cite{Chohetero}. The Kerr-Schild formalism is also used to relate the double copy idea with the usual DFT equations \cite{Leekerr}, and with T-duality \cite{Angus}.  In this work, we do not use the Kerr-Schild ansatz for heterotic DFT.  \\

The starting point of our work is inspired by the results of \cite{expform}, which states that the amplitudes for a Yang-Mills-Einstein (YME) theory coupled to a dilaton and two-form field can be obtained by the double copy of pure Yang-Mills theory and Yang-Mills+$\phi^3$ ($YM+\phi^3$) theory. Following this work, we use this $YM+\phi^3$ theory introduced in \cite{expform} as a starting point and take its double copy with a pure Yang-Mills theory. For simplicity, we restrict ourselves to the Abelian sector of heterotic DFT. The action we constructed by a double copy procedure exactly matches with the Abelian sector of heterotic DFT action up to cubic order.\\

We believe that our results can help us to understand the double copy relations at the Lagrangian level. Although the double copy relations at the level of amplitudes are well established, it is hard to say the same thing for double copy relations at the Lagrangian level. In that sense, having a double copy formulation for any theory at the Lagrangian level is helpful to get more insights about the nature of double copy. Moreover, having a double copy formulation of heterotic DFT can help us to understand the way which goes to the double copy construction of supergravity theories.

\section{\label{sec1}DFT as Double Copy of Yang-Mills Theory}
In this section, we review the usual construction of DFT action by a double copy procedure. The action for the pure Yang-Mills theory in $D$-dimensions is given by,
\begin{equation}
    S_{\mathrm{YM}}=-\frac{1}{4} \int d^D x \kappa_{a b} F^{\mu \nu a} F_{\mu \nu}^b,
\label{ymaction}
\end{equation}
where the field strength is defined as
\begin{equation}
    F_{\mu \nu}{ }^a=\partial_\mu A_\nu{ }^a-\partial_\nu A_\mu{ }^a+g_{\mathrm{YM}} f^a{ }_{b c} A_\mu{ }^b A_\nu{ }^c,
\label{fieldstrength}
\end{equation}
with $g_{YM}$ is the coupling constant and $f^a{}_{bc}$ are the structure constants. Expanding up to quadratic order, integrating by parts and passing to the momentum space by 
\begin{equation}
    A_\mu^a(k) \equiv \frac{1}{(2 \pi)^{D / 2}} \int d^D x A_\mu^a(x) e^{i k x},
\end{equation}
the action (\ref{ymaction}) takes the form
\begin{equation}
    S_{\mathrm{YM}}^{(2)}=-\frac{1}{2} \int_k \kappa_{a b} k^2 \Pi^{\mu \nu}(k) A_\mu^a(-k) A_\nu^b(k),
\label{eqn1}
\end{equation}
where $\int_k:=\int d^D k$ and the projector is defined as
\begin{equation}
    \Pi^{\mu \nu}(k) \equiv \eta^{\mu \nu}-\frac{k^\mu k^\nu}{k^2}.
\label{projector}
\end{equation}
Following \cite{jaram}, one can introduce the double copy maps,
    \begin{equation}
    A_\mu{ }^a(k) \rightarrow e_{\mu \bar{\mu}}(k, \bar{k}) \quad \text{and} \quad \kappa_{a b} \longrightarrow \frac{1}{2} \bar{\Pi}^{\bar{\mu} \bar{\nu}}(\bar{k}).
\label{dcmap}
\end{equation}
Inserting these into the (\ref{eqn1}) and introducing the dilaton field by
\begin{equation}
    \phi=\frac{1}{k^2} k^\mu \bar{k}^{\bar{\nu}} e_{\mu \bar{\nu}},
\end{equation}
the action (\ref{eqn1}) becomes
\begin{equation}
    \begin{aligned}
S_{\mathrm{DC}}^{(2)}=  -\frac{1}{4} \int_{k, \bar{k}} & \left(k^2 e^{\mu \bar{\nu}} e_{\mu \bar{\nu}}-k^\mu k^\rho e_{\mu \bar{\nu}} e_\rho^{\bar{\nu}}-\bar{k}^{\bar{\nu}} \bar{k}^{\bar{\sigma}} e_{\mu \bar{\nu}} e^\mu \bar{\sigma}\right. \\
& \left.-k^2 \phi^2+2 \phi k^\mu \bar{k}^{\bar{\nu}} e_{\mu \bar{\nu}}\right) .
\end{aligned}
\end{equation}
Passing to the position space, this action reads,
\begin{equation}
    \begin{aligned}
S_{\mathrm{DC}}^{(2)}= & \frac{1}{4} \int d^D x d^D \bar{x}\left(e^{\mu \bar{\nu}} \square e_{\mu \bar{\nu}}+\partial^\mu e_{\mu \bar{\nu}} \partial^\rho e_\rho^{\bar{\nu}}\right. \\
& \left.+\bar{\partial}^{\bar{\nu}} e_{\mu \bar{\nu}} \bar{\partial}^{\bar{\sigma}} e^\mu{ }_{\bar{\sigma}}-\phi \square \phi+2 \phi \partial^\mu \bar{\partial}^{\bar{\nu}} e_{\mu \bar{\nu}}\right),
\end{aligned}
\end{equation}
which defines precisely the standard quadratic double field theory action. Similarly, one can start with the cubic part of the Yang-Mills action,
\begin{equation}
    S_{\mathrm{YM}}^{(3)}=-g_{\mathrm{YM}} \int d^D x f_{a b c} \partial^\mu A^{\nu a} A_\mu^b A_\nu^c,
\label{cubic}
\end{equation}
and use double copy procedure to obtain the cubic part of the DFT action. Passing to the momentum space and writing the action more symmetrically, this cubic part (\ref{cubic}) reads
\begin{equation}
    \begin{aligned}
S_{\mathrm{YM}}^{(3)}= & -\frac{i g_{\mathrm{YM}}}{6(2 \pi)^{D / 2}} \int_{k_1, k_2, k_3} \delta\left(k_1+k_2+k_3\right) \\
& \times f_{a b c} \Pi^{\mu \nu \rho}\left(k_1, k_2, k_3\right) A_{1 \mu}{ }^a A_{2 \nu}{ }^b A_{3 \rho}{ }^c,
\end{aligned}
\label{cubicmom}
\end{equation}
where the projector is defined as
\begin{equation}
    \Pi^{\mu \nu \rho}\left(k_1, k_2, k_3\right) \equiv \eta^{\mu \nu} k_{12}^\rho+\eta^{\nu \rho} k_{23}^\mu+\eta^{\rho \mu} k_{31}^\nu,
\end{equation}
with $k_{i j} \equiv k_i-k_j$. This projector satisfies the same antisymmetry properties with the structure constant $f_{abc}$. Introducing the double copy map,
\begin{equation}
    f_{a b c} \rightarrow \frac{i}{4} \bar{\Pi}^{\bar{\mu} \bar{\nu} \bar{\rho}}\left(\bar{k}_1, \bar{k}_2, \bar{k}_3\right),
\label{fdcmap}
\end{equation} 
inserting this to (\ref{cubicmom}), passing to the position space and integrating by parts one obtains
\begin{equation}
    \begin{aligned}
S_{\mathrm{DC}}^{(3)}= & \frac{1}{8} \int d^D x d^D \bar{x} e_{\mu \bar{\mu}}\left[2 \partial^\mu e_{\rho \bar{\rho}} \bar{\partial}^{\bar{\mu}} e^{\rho \bar{\rho}}-2 \partial^\mu e_{\nu \bar{\rho}} \bar{\partial}^{\bar{\rho}} e^{\nu \bar{\mu}}\right. \\
& \left.-2 \partial^\rho e^{\mu \bar{\rho}} \partial^{\bar{\mu}} e_{\rho \bar{\rho}}+\partial^\rho e_{\rho \bar{\rho}} \bar{\partial}^{\bar{\rho}} e^{\mu \bar{\mu}}+\bar{\partial}_{\bar{\rho}} e^{\mu \bar{\rho}} \partial_\rho e^{\rho \bar{\mu}}\right] .
\end{aligned}
\label{cubicdc}
\end{equation}
As it was shown in \cite{jaram}, this cubic action (\ref{cubicdc}) agrees with the cubic DFT upon imposing a gauge-fixing condition and integrating out the dilaton. \\
\section{Heterotic DFT Formulation}
In this section, we review the DFT formulation for the abelian subsector of the low-energy theory of the heterotic string following \cite{Hohmhetero}. Let me start with the generalized metric formulation of usual DFT action. Introducing the generalized metric on the doubled space with coordinates $X^M=\left(\tilde{x}_i, x^i\right)$,
\begin{equation}
    \mathcal{H}^{M N}=\left(\begin{array}{cc}
g_{i j}-b_{i k} g^{k l} b_{l j} & b_{i k} g^{k j} \\
-g^{i k} b_{k j} & g^{i j}
\end{array}\right),
\end{equation}
one can write the DFT action as \cite{hohmdft},
\begin{equation}
    \begin{aligned}
S=\int_{x,\tilde{x}} & e^{-2 d}(  \frac{1}{8} \mathcal{H}^{M N} \partial_M \mathcal{H}^{K L} \partial_N \mathcal{H}_{K L}+4 \mathcal{H}^{M N} \partial_M d \partial_N d \\
& -\frac{1}{2} \mathcal{H}^{M N} \partial_N \mathcal{H}^{K L} \partial_L \mathcal{H}_{M K}-2 \partial_M d \partial_N \mathcal{H}^{M N}),
\end{aligned}
\label{dftaction}
\end{equation}
with derivatives $\partial_M=(\tilde{\partial}^i, \partial_i)$. In this action, the $O(D,D)$ invariant dilaton $d$ defined by $e^{-2 d}=\sqrt{g} e^{-2 \phi}$, the indices are raised and lowered by $O(D,D)$ invariant metric,
\begin{equation}
    \eta_{M N}=\left(\begin{array}{ll}
0 & 1 \\
1 & 0
\end{array}\right).
\end{equation} 
As it was shown in \cite{Hohmhetero}, to construct the DFT formulation of heterotic strings, one can start with enlarging the coordinates as,
\begin{equation}
    X^M=\left(\tilde{x}_i, x^i, y^\alpha\right),
\end{equation}
which transforms as a fundamental $O(D,D+n)$ vector \cite{Hohmhetero}, and $y^\alpha$ correspond to the coordinates for the internal space of gauge group. Then, the $O(D,D+n)$ invariant metric takes the form
\begin{equation}
    \eta_{M N}=\left(\begin{array}{ccc}
\eta^{i j} & \eta_j{ }_j & \eta^i{ }_\beta \\
\eta_i{ }^j & \eta_{i j} & \eta_{i \beta} \\
\eta_\alpha{ }^j & \eta_{\alpha j} & \eta_{\alpha \beta}
\end{array}\right)=\left(\begin{array}{ccc}
0 & 1 & 0 \\
1 & 0 & 0 \\
0 & 0 & \kappa
\end{array}\right),
\label{heteroinvmetric}
\end{equation}
where $\kappa$ denote the matrix corresponding to the Cartan-Killing metric of the gauge group \cite{Hohmhetero}. Next, one can introduce the extended form of the generalized metric which transforms covariantly under $O(D,D+n)$ transformations. In \cite{Hohmhetero}, this extended form is given as,
\begin{equation}
\begin{aligned}
    & \mathcal{H}_{M N} = \\
&\left(\begin{array}{ccc}
g^{i j} & -g^{i k} c_{k j} & -g^{i k} A_{k \beta} \\
-g^{j k} c_{k i} & g_{i j}+c_{k i} g^{k l} c_{l j}+A_i^\gamma A_{j \gamma} & c_{k i} g^{k l} A_{l \beta}+A_{i \beta} \\
-g^{j k} A_{k \alpha} & c_{k j} g^{k l} A_{l \alpha}+A_{j \alpha} & \kappa_{\alpha \beta}+A_{k \alpha} g^{k l} A_{l \beta}
\end{array}\right),
\end{aligned}
\label{heterogenmetric}
\end{equation}
where,
\begin{equation}
    c_{i j}=b_{i j}+\frac{1}{2} A_i^\alpha A_{j \alpha}.
\label{cdef}
\end{equation}
Inserting these extended objects (\ref{heteroinvmetric}), (\ref{heterogenmetric}) together with $\partial_M=(\tilde{\partial}^i, \partial_i, \partial_\alpha)$ into the action (\ref{dftaction}) and setting $\tilde{\partial}^i=\partial_\alpha=0$, one obtains (for the abelian case),
\begin{equation}
    \begin{aligned}
S=\int & d^D x e^{-2 d}\left(\frac{1}{4} g^{i j} \partial_i g^{k l} \partial_j g_{k l}-\frac{1}{2} g^{i j} \partial_j g^{k l} \partial_l g_{i k}\right. \\
&\left.-2 \partial_i d \partial_j g^{i j}+4 g^{i j} \partial_i d \partial_j d-\frac{1}{12} \hat{H}^2-\frac{1}{4} F_{i j \alpha} F^{i j \alpha}\right),
\end{aligned}
\label{heterodftlimitaction1}
\end{equation}
where $\hat{H}_{i j k}=3\left(\partial_{[i} b_{j k]}-\kappa_{\alpha \beta} A_{[i}^\alpha(\partial_j A_{k]}^\beta)\right)$. Up to boundary terms, the abelian sector of this action (\ref{heterodftlimitaction1}) coincides with the abelian sector of the low-energy limit of heterotic string action \cite{Hohmhetero}, which is
\begin{equation}
    S=\int d^D x \sqrt{g} e^{-2 \phi}(R+4(\partial \phi)^2-\frac{1}{12} \hat{H}^2-\frac{1}{4} F^{i j \alpha} F_{i j \alpha}).
\label{heterodftlimitaction2}
\end{equation}
For the comparison with the results of the following sections, one can write the action (\ref{heterodftlimitaction1}) around Minkowski space, $g_{ij} = \eta_{ij}+h_{ij}$, in the form,
\begin{equation}
        S=\int d^D x \left(\mathcal{A}_1+\mathcal{A}_2+\mathcal{A}_3+\mathcal{A}_4+\mathcal{A}_5 \right),
\label{heterodftlimitaction}
\end{equation}
where,
\begin{align}
    &\mathcal{A}_1 \equiv +\frac{1}{4} \eta^{i j} \partial_i g^{k l} \partial_j g_{k l}-\frac{1}{2} \eta^{i j} \partial_j g^{k l} \partial_l g_{i k}-2 \partial_i d \partial_j h^{i j} \nonumber \\
    & \quad \quad \quad +4 \eta^{i j} \partial_i d \partial_j d - \frac{1}{12}H^{ijk}H_{ijk}, \label{usualdftterms}\\
    &\mathcal{A}_2 \equiv +\frac{1}{4}(\partial_\sigma  e_{\mu \beta})  \bigg( (\partial^\sigma A^{\mu})A^\beta-(\partial^\sigma A^{\beta})A^\mu+(\partial^\beta A^{\sigma})A^\mu  \nonumber \\
    & \quad \quad \quad \quad \quad -(\partial^\beta A^{\mu})A^\sigma+(\partial^\mu A^{\beta})A^\sigma-(\partial^\mu A^{\sigma})A^\beta \bigg), \label{eAAterms}\\
    &\mathcal{A}_3 \equiv -\frac{1}{8}(\partial_\sigma A_\mu)A_\beta \bigg((\partial^\sigma A^{\mu})A^\beta-(\partial^\sigma A^{\beta})A^\mu \nonumber  \\
    &\quad +(\partial^\beta A^{\sigma})A^\mu -(\partial^\beta A^{\mu})A^\sigma+(\partial^\mu A^{\beta})A^\sigma-(\partial^\mu A^{\sigma})A^\beta \bigg), \label{AAAAterms}\\
    &\mathcal{A}_4 \equiv -\frac{1}{4} \kappa_{ab} F^{i j a} F_{i j}{}^b, \label{ymterms} \\ 
    &\mathcal{A}_5 \equiv \quad \textit{Higher order terms } \label{higherorderterms},
\end{align}
with $H_{ijk}$ is the field strength for the two-form field $b_{ij}$.
\section{Double Copy Formulation for Heterotic DFT}
In \cite{expform}, it has been shown that the scattering amplitudes of Yang-Mills-Einstein theory coupled the dilaton and two-form field $B^{\mu \nu}$ can be obtained from the double copy of (Yang-Mills+$\phi^3$) theory and pure Yang-Mills theory. The Lagrangian for the (Yang-Mills+$\phi^3$) is given in \cite{expform} as,
\begin{equation}
    \begin{aligned}
\mathcal{L}_{\mathrm{YM}+\phi^3}= & -\frac{1}{4} \kappa_{ab}F_{\mu \nu}{}^{a} F^{\mu \nu b}+\frac{1}{2}\kappa_{ab} \kappa_{AB}D_\mu \phi^{Aa}D^\mu \phi^{B{b}} \\
&-\frac{g^2}{4} \kappa_{AC}\kappa_{BD}\kappa^{ef} f_{{a}{b} e}{} f_{{f} {c}{d}} \phi^{A {a}} \phi^{B {b}} \phi^{C {c}} \phi^{D {d}} \\
& +\frac{1}{3!} \lambda g F_{A B C} f_{{a} {b} {c}} \phi^{A {a}} \phi^{B {b}} \phi^{C {c}},
\end{aligned}
\label{phitheory}
\end{equation}
where the $a,b,c,d$ indices run over the adjoint representation of the gauge group and scalar fields carry additional indices $A, B, C=1,2, \ldots, n$. Field strength and covariant derivative are defined as \cite{expform}
\begin{equation}
    \begin{aligned}
F_{\mu \nu}^{{a}} & =\partial_\mu A_\nu{}^{{a}}-\partial_\nu A_\mu{}^{{a}}+g f^{a}{}_{{b}{c}} A_\mu{}^{{b}} A_\nu{}^{{c}}, \\
D_\mu \phi^{A{a}} & =\partial_\mu \phi^{A {a}}+g f^{a}{}_{{b}{c}} A_\mu{}^{{b}} \phi^{A {c}} .
\end{aligned}
\end{equation}
The Lagrangian for the usual Yang-Mills action is,
\begin{equation}
    \mathcal{L}_{\mathrm{YM}}=-\frac{1}{4} \kappa_{a b} {F}^{\mu \nu a} {F}_{\mu \nu}^b.
\label{tildeymaction}
\end{equation}
Now, we use these two Lagrangians (\ref{phitheory}) and (\ref{tildeymaction}) to construct the Abelian subsector of heterotic DFT action. To obtain the Abelian subsector, let us set the structure constants as $F_{ABC}=0$ while $f_{abc}$ are nonzero. After this choice, one can expand the Lagrangian (\ref{phitheory}) explicitly and separate it into five parts as,
\begin{equation}
    \begin{aligned}
        \mathcal{L}_{\phi1} &=  -\frac{1}{4} \kappa_{ab}F_{\mu \nu}{}^{a} F^{\mu \nu b},\\
        \mathcal{L}_{\phi2} &=+\frac{1}{2}\kappa_{ab} \kappa_{AB}\,\eta^{\mu \nu}(\partial_{\mu} \phi^{Aa})(\partial_{\nu} \phi^{Bb}), \\
        \mathcal{L}_{\phi3} &=+g \, \kappa_{AB} \, \eta^{\mu \nu} f_{abc} (\partial_\mu \phi^{Aa})(A_{\nu}{}^b \phi^{Bc}),\\
        \mathcal{L}_{\phi4} &=+\frac{1}{2}g^2\kappa^{ab} \kappa_{AB}\,\eta^{\mu \nu}(f_{acd} A_\mu{}^c \phi^{Ad})(f_{bef} A_\nu{}^e \phi^{Bf}),\\
        \mathcal{L}_{\phi5} &=-\frac{g^2}{4} \kappa_{AC}\kappa_{BD}\kappa^{ef} f_{{a}{b} e}{} f_{{f} {c}{d}} \phi^{A {a}} \phi^{B {b}} \phi^{C {c}} \phi^{D {d}}.
    \end{aligned}
\label{fivelag}
\end{equation}
We start with the known double copy maps in the literature and determine the other ones to find a match with (\ref{heterodftlimitaction}). Note that while applying double copy maps on these Lagrangians, we replace the color information in these Lagrangians with the kinematic factors of the action (\ref{tildeymaction}). In that sense, the double copy structure is applied on the two actions (\ref{phitheory}) and (\ref{tildeymaction}). That should not be confused with double copying $(YM+\phi^3)$ theory with itself. Such a construction can lead another theory, which is out of the scope of this paper. \\

Passing to the momentum space, and introducing double copy maps (\ref{dcmap}), (\ref{fdcmap}), $\mathcal{L}_{\phi1}$ leads to usual DFT action up to cubic order \cite{jaram}, the terms in (\ref{usualdftterms}) and corresponding cubic terms in (\ref{higherorderterms}). \\

Now, consider the second piece, $\mathcal{L}_{\phi2}$. Passing to the momentum space, it reads
\begin{equation}
    S_{\phi2} = -\frac{1}{2} \int_{k_1,k_2} \delta(k_1+k_2)\kappa_{ab} \kappa_{AB} (\eta^{mn}k_{1m}k_{2n}) \phi_1^{Aa} \phi_2^{Bb},
\label{secondactioninmomspace}
\end{equation}
with the notation $\phi_i^{Aa}=\phi^{Aa}(k_i)$. The corresponding Lagrangian in momentum space can be structurally written as
\begin{equation}
    \mathcal{L}_{\phi2} \sim -\frac{1}{2}\kappa_{ab} \kappa_{AB} (\eta^{mn}k_{1m}k_{2n}) \phi_1^{Aa} \phi_2^{Bb}.
\end{equation}
At this point, we divide this Lagrangian into two and write, 
\begin{equation}
    \begin{aligned}
        &\mathcal{L}_{\phi21} = -\frac{1}{4}\kappa_{ab} \kappa_{AB} (\eta^{mn}k_{1m}k_{2n}) \phi_1^{Aa} \phi_2^{Bb}, \\
        &\mathcal{L}_{\phi22} = -\frac{1}{4}\kappa_{ab} \kappa_{AB} (\eta^{mn}k_{1m}k_{2n}) \phi_1^{Aa} \phi_2^{Bb}.
    \end{aligned}
\end{equation}
The purpose of separating Lagrangian into two parts is to define double copy maps in a symmetric way in barred and unbarred indices, which will be clear in a moment. \\

Let us start with $\mathcal{L}_{\phi21}$. The double copy map for $\kappa_{ab}$ is already introduced in (\ref{dcmap}). However, in this case the role of $\kappa_{ab}$ is to contract $\phi^{Aa}$ fields instead of contracting gauge fields $A_{\mu}{}^a$. So, it is possible that its double copy differs from (\ref{dcmap}), at least for the coefficient. We want to keep all the possibilities in the construction, so for now we define the double copy map with an arbitrary coefficient, as,
\begin{equation}
    \kappa_{ab} \equiv a_1 \left(\bar{\eta}^{\alpha \beta}-\frac{\bar{k}_1^\beta \bar{k}_2^\alpha }{\bar{\eta}^{pq}\bar{k}_{1p}\bar{k}_{2q}} \right)\equiv a_1 \bar{\Pi}^{\alpha \beta},
\label{kappadc}
\end{equation}
while the double copy maps for $\kappa_{AB}$ and $\phi^{Aa}$ are not yet known. All we can guess is that the indices $A,B$ should be mapped to the indices $a,b$; while the indices $a,b$ should be mapped to the spacetime indices as in (\ref{dcmap}). After a careful analysis, we introduce ansatz for them as,
\begin{equation}
\begin{aligned}
    &\kappa_{AB} \rightarrow a_2 \kappa_{ab}, \\
    &\phi^{Aa}(k_i) \rightarrow a_3\bar{A}_\alpha{}^a(k_i,\bar{k}_i)+a_4e_{\mu \alpha}A^{\mu a}(k_i,\bar{k}_i),
\end{aligned}
\label{ansatzdc}
\end{equation}
where the constants $a_i$ will be determined. Using these double copy maps (\ref{kappadc}), (\ref{ansatzdc}), and also integrating in the dual coordinates one gets
\begin{equation}
    S'_{\phi21} = \int_{K_1,K_2} \delta(K_1+K_2)\left(\mathcal{L}_{\phi211} + \mathcal{L}_{\phi212} + \mathcal{L}_{\phi213} \right),
\label{phi2dcactionimplicitform}
\end{equation}
where,
\begin{equation}
    \begin{aligned}
        &\mathcal{L}_{\phi211} \equiv -\frac{a_1a_2a_3^2}{4} (k_{1m}k_{2}^m) \kappa_{ab} \bar{\Pi}^{\alpha \beta}\bar{A}_\alpha{}^a(k_1,\bar{k}_1)\bar{A}_\beta{}^b(k_2,\bar{k}_2),\\
        &\mathcal{L}_{\phi212} \equiv -\frac{a_1a_2a_3a_4}{2} (k_{1m}k_{2}^m) \kappa_{ab} \bar{\Pi}^{\alpha \beta}\bar{A}_\alpha{}^a(k_1,\bar{k}_1)\bar{f}_\beta{}^b(k_2,\bar{k}_2),\\
        &\mathcal{L}_{\phi213} \equiv -\frac{a_1a_2a_4^2}{4} (k_{1m}k_{2}^m) \kappa_{ab} \bar{\Pi}^{\alpha \beta}\bar{f}_\alpha{}^a(k_1,\bar{k}_1)\bar{f}_\beta{}^b(k_2,\bar{k}_2),
    \end{aligned}
\label{phi2lags}
\end{equation}
with the notation $\bar{f}_\alpha{}^a(k_1,\bar{k}_1) = e_{\mu \alpha}A^{\mu a}$. \\

Now we continue with $\mathcal{L}_{\phi 22}$, with the corresponding double copy maps,
\begin{equation}
    \begin{aligned}
        &\kappa_{ab} \rightarrow b_1 \Pi^{\mu \nu} \\
        &\kappa_{AB} \rightarrow a_2 \kappa_{ab}\\
        &\phi^{Aa} \rightarrow b_3A_\mu{}^a(k_i,\bar{k}_i)+b_4e_{\mu \alpha}\bar{A}^{\alpha a}(k_i,\bar{k}_i).
    \end{aligned}
\label{quadsymdcmaps}
\end{equation}
At this point, it is important to justify our choices for the coefficients in (\ref{quadsymdcmaps}), by comparing them with the choices in (\ref{kappadc}) and (\ref{ansatzdc}). In (\ref{kappadc}), $\kappa_{ab}$ is mapped to the $\bar{\Pi}^{\alpha \beta}$ which has barred indices and in (\ref{quadsymdcmaps}) $\kappa_{ab}$ is mapped to the $\Pi^{\mu \nu}$ with unbarred indices. In these two cases, $\kappa_{ab}$ is mapped to the objects with different index structure, so we use different coefficients for these two cases. Similarly, we use different coefficients for the double copy maps of $\phi^{Aa}$ in (\ref{ansatzdc}) and (\ref{quadsymdcmaps}). On the other hand, $\kappa_{AB}$ is mapped to the $\kappa_{ab}$ in both (\ref{ansatzdc}) and (\ref{quadsymdcmaps}), so we choose their coefficients same. Using these double copy maps (\ref{quadsymdcmaps}), we obtain,
\begin{equation}
    S'_{\phi22} = \int_{K_1,K_2} \delta(K_1+K_2)\left(\mathcal{L}_{\phi221} + \mathcal{L}_{\phi222} + \mathcal{L}_{\phi223} \right),
\label{phi2dcactionimplicitform2}
\end{equation}
where,
\begin{equation}
    \begin{aligned}
        &\mathcal{L}_{\phi221} \equiv -\frac{b_1a_2b_3^2}{4} (k_{1m}k_{2}^m) \kappa_{ab} {\Pi}^{\mu \nu}A_\mu{}^a(k_1,\bar{k}_1)A_\nu{}^b(k_2,\bar{k}_2),\\
        &\mathcal{L}_{\phi222} \equiv -\frac{b_1a_2b_3b_4}{2} (k_{1m}k_{2}^m) \kappa_{ab} {\Pi}^{\mu \nu}A_\mu{}^a(k_1,\bar{k}_1)f_\nu{}^b(k_2,\bar{k}_2),\\
        &\mathcal{L}_{\phi223} \equiv -\frac{b_1a_2b_4^2}{4} (k_{1m}k_{2}^m) \kappa_{ab} {\Pi}^{\mu \nu}f_\mu{}^a(k_1,\bar{k}_1)f_\nu{}^b(k_2,\bar{k}_2),
    \end{aligned}
\label{phi22lags}
\end{equation}
with the notation $f_\mu{}^a(k_1,\bar{k}_1) = e_{\mu \alpha}\bar{A}^{\alpha a}$. \\

So, starting with the Lagrangian $\mathcal{L}_{\phi2}$ in (\ref{fivelag}) and using double copy procedure, we obtain (\ref{phi2lags}) and (\ref{phi22lags}). To fix the coefficients $a_i$ and $b_i$, we can compare these terms with (\ref{heterodftlimitaction2}).
Consider the terms $\mathcal{L}_{\phi211}$ and $\mathcal{L}_{\phi221}$, and their corresponding actions
\begin{equation}
\begin{aligned}
    &S'_{\phi211} = \frac{-a_1a_2a_3^2}{4} \int_{K_i} \delta(K_i)(k_{1m}k_{2}^m) \kappa_{ab} \bar{\Pi}^{\alpha \beta}\bar{A}_{1\alpha}{}^a\bar{A}_{2\beta}{}^b, \\
    &S'_{\phi221}= \frac{-b_1a_2b_3^2}{4}\int_{K_i} \delta(K_i)(k_{1m}k_{2}^m) \kappa_{ab}\Pi^{\mu \nu}A_{1\mu}{}^aA_{2\nu}{}^b,
\end{aligned}
\label{sprimephi211221}
\end{equation}
with the notation,
\begin{equation}
    K_i \equiv (k_i, \bar{k}_i), \quad \delta(K_i) \equiv \delta(K_1+K_2), \quad \int_{K_i} \equiv \int dK_1dK_2.
\end{equation}
Note that, in the construction above the derivative operators with bars on top of them can be thought as corresponding to the right moving coordinates, while the usual derivative operators corresponds to the left moving coordinates. In DFT language, this can be written as
\begin{equation}
    \partial^\mu \equiv D^\mu \quad \text{and} \quad \bar{\partial}^\alpha \equiv \bar{D}^\alpha,
\end{equation}
and then turning off the dual coordinates means that 
\begin{equation}
    D^\mu = \bar{D}^\mu \quad \text{or} \quad \partial^\mu = \bar{\partial}^\mu.
\end{equation}
Turning of the dual coordinates, the sum of the actions in (\ref{sprimephi211221}) becomes
\begin{equation}
    \mathcal{B}_1 = \frac{-C_1}{4} \int_{k_i} \delta(k_i)(k_{1m}k_{2}^m) \kappa_{ab} {\Pi}^{\alpha \beta}{A}_{1\alpha}{}^a{A}_{2\beta}{}^b,
\label{phi21sym}
\end{equation}
where $C_1 = a_1a_2a_3^2+b_1a_2b_3^2$. Similarly, in momentum space, the Abelian sector of (\ref{ymterms}) reads
\begin{equation}
    \mathcal{A}_4 = \frac{1}{2}\int_{k_i} \delta(k_i)(k_{1m}k_{2}^m) \kappa_{ab} {\Pi}^{\alpha \beta}{A}_{1\alpha}{}^a{A}_{2\beta}{}^b.
\label{a4}
\end{equation}
Equating (\ref{phi21sym}) and (\ref{a4}) gives,
\begin{equation}
    C_1 = a_1a_2a_3^2+b_1a_2b_3^2 = -2,
\label{firstconstraint}
\end{equation}
which is our first constraint on the undetermined coefficients. \\

Now, we can continue with $\mathcal{L}_{\phi212}$ and $\mathcal{L}_{\phi222}$. Passing to the position space, these Lagrangian terms give
\begin{equation}
    \begin{aligned}
        \mathcal{L}_{\phi212} = \frac{a_1a_2a_3a_4}{2} \kappa_{ab} \bigg( &(\partial_m \bar{A}^{\beta a}) \partial^m(e_{\mu \beta}A^{\mu b}) \\
        &-(\bar{\partial}^\beta \bar{A}_\alpha{}^a)\bar{\partial}^\alpha(e_{\mu \beta}A^{\mu b}) \bigg),
    \end{aligned}
    \label{lphi212}
\end{equation}
\begin{equation}
    \begin{aligned}
        \mathcal{L}_{\phi222} = \frac{b_1a_2b_3b_4}{2} \kappa_{ab} \bigg( &(\partial_m {A}^{\nu a}) \partial^m(e_{\nu \beta}\bar{A}^{\beta b}) \\
        &-({\partial}^\nu {A}_{\mu}{}^a){\partial}^\mu(e_{\nu \beta}\bar{A}^{\beta b}) \bigg).
    \end{aligned}
    \label{lphi222}
\end{equation}
Turning off the dual coordinates and summing the terms (\ref{lphi212}) and (\ref{lphi222}) one gets,
\begin{equation}
\begin{aligned}
    \mathcal{B}_2 = &+\frac{a_1a_2a_3a_4}{2} \kappa_{ab} (\partial_m e_{\mu \beta})(\partial^m A^{\beta a})A^{\mu b} \\
    &-\frac{a_1a_2a_3a_4}{2} \kappa_{ab} (\partial_\alpha e_{\mu \beta})(\partial^\beta A^{\alpha a})A^{\mu b},
\end{aligned}
\label{b2}
\end{equation}
\begin{equation}
    \begin{aligned}
        \mathcal{B}_3 = &+\frac{b_1a_2b_3b_4}{2} \kappa_{ab}(\partial_m e_{\nu \beta})(\partial^m A^{\nu a})A^{\beta b}\\
        &-\frac{b_1a_2b_3b_4}{2} \kappa_{ab}(\partial_\mu e_{\nu \beta})(\partial^\nu A^{\mu a})A^{\beta b},
    \end{aligned}
    \label{b3}
\end{equation}
\begin{equation}
    \begin{aligned}
        \mathcal{B}_4 =&+\frac{a_1a_2a_3a_4}{2} \kappa_{ab} e_{\mu \beta}(\partial_m A^{\beta a})(\partial^m A^{\mu b})\\
        &+\frac{b_1a_2b_3b_4}{2}\kappa_{ab} e_{\nu \beta} (\partial_m A^{\beta a})(\partial^mA^{\nu b}),
    \end{aligned}
    \label{b4}
\end{equation}
\begin{equation}
    \begin{aligned}
        \mathcal{B}_5 =&-\frac{a_1a_2a_3a_4}{2} \kappa_{ab} e_{\mu \beta}(\partial^\beta A_\alpha{}^a)(\partial^\alpha A^{\mu b})\\
        &-\frac{b_1a_2b_3b_4}{2}\kappa_{ab} e_{\nu \beta}(\partial^\nu A_\mu{}^a)(\partial^\mu A^{\beta b}).
    \end{aligned}
    \label{b5}
\end{equation}
Now, we compare these terms with the terms of (\ref{heterodftlimitaction1}). When the coefficients satisfy
\begin{equation}
    a_1a_2a_3a_4 = -\frac{1}{2} \quad \text{and} \quad b_1a_2b_3b_4=+\frac{1}{2},
\label{secondconstraint}
\end{equation}
$\mathcal{B}_2$ gives second and third terms of $\mathcal{A}_2$ in (\ref{eAAterms}), and $\mathcal{B}_3$ gives the first and sixth terms of $\mathcal{A}_2$. For the choice (\ref{secondconstraint}), $\mathcal{B}_4$ and $\mathcal{B}_5$ becomes,
\begin{equation}
    \begin{aligned}
        \mathcal{B}_4 =& 0 \\
        \mathcal{B}_5 = & \frac{1}{4} \kappa_{ab} e_{\mu \beta} \left( (\partial^\beta A_\alpha{}^a)(\partial^\alpha A^{\mu b}) - (\partial^\mu A_\nu{}^a)(\partial^\nu A^{\beta b})\right).     
    \end{aligned}
\label{b5extra}
\end{equation}
Let us summarize what we have done so far. Using double copy procedure on $\mathcal{L}_{\phi1}$ and $\mathcal{L}_{\phi2}$ from (\ref{fivelag}), we obtain $\mathcal{A}_1$ (\ref{usualdftterms}), $\mathcal{A}_4$ (\ref{ymterms}) and four of the six terms in $\mathcal{A}_2$ (\ref{eAAterms}). We also obtain some extra terms $\mathcal{B}_5$, given in (\ref{b5extra}). At this point, we can continue by calculating the contributions from $\mathcal{L}_{\phi213}$ from (\ref{phi2lags}) and $\mathcal{L}_{\phi223}$ from (\ref{phi22lags}). However, they give fourth order terms, structurally $eeAA$ terms, and these are out of the scope of this work. \\

Hence, we continue by calculating the contributions from $\mathcal{L}_{\phi3}$ from (\ref{fivelag}), with the corresponding action in momentum space,
\begin{equation}
    S_{\phi3} = \frac{-ig}{(2\pi)^{D/2}} \int_{k} \kappa_{AB} f_{abc} \delta(k_1+k_2+k_3) \, k^{\mu}_1 \phi_1^{Aa} A_{2\mu}{}^b\phi_3^{Bc}.
\label{sphi3}
\end{equation}
The corresponding Lagrangian can be structurally written as,
\begin{equation}
    \mathcal{L}_{\phi3} \sim \frac{-i}{2} \kappa_{AB} f_{abc}\, k^{\mu}_1 \phi_1^{Aa} A_{2\mu}{}^b\phi_3^{Bc}
\end{equation}
where we also used $g \rightarrow \frac{1}{2}$. We also separate this Lagrangian into two parts, to write the final action in a symmetric way with respect to barred and unbarred indices. So we have
\begin{equation}
    \begin{aligned}
        \mathcal{L}_{\phi31} = &\frac{-i}{4} \kappa_{AB} f_{abc} \, k^{\mu}_1 \phi_1^{Aa} A_{2\mu}{}^b\phi_3^{Bc},\\
        \mathcal{L}_{\phi32} = &\frac{-i}{4} \kappa_{AB} f_{abc} \, \bar{k}^{\alpha}_1 \phi_1^{Aa} \bar{A}_{2\alpha}{}^b\phi_3^{Bc}.
    \end{aligned}
\label{lphi3lags}
\end{equation}
Let us start with $\mathcal{L}_{\phi31}$ and introduce the double copy maps
\begin{equation}
    \begin{aligned}
        &\kappa_{AB} \rightarrow a_2 \kappa_{ab},\\
        &f_{abc} \rightarrow c_2 \bar{\Pi}^{\alpha \beta \sigma}(\bar{k}_1,\bar{k}_2,\bar{k}_3),\\
        &\phi^{Aa}(k_i) \rightarrow a_3\bar{A}_\alpha{}^a(k_i, \bar{k}_i)+a_4e_{\nu \alpha}A^{\nu b},\\
        &A_\mu{}^b(k_i) \rightarrow e_{\mu \beta}(k_i, \bar{k}_i).
    \end{aligned}
\label{cubicdcterms}
\end{equation}
The coefficients for the double maps of $\kappa_{AB}$, $\phi^{Aa}$ and $A_\mu{}^b$ are already fixed by the previous results. For the double copy map of $f_{abc}$ we use an arbitrary coefficient, because in that case this structure constant contract the indices of $\phi^{Aa}$, $A_\mu{}^b$ and $\phi^{Bc}$.  Inserting these maps into $\mathcal{L}_{\phi31}$ one gets,
\begin{equation}
    \begin{aligned}
        &\mathcal{L}_{\phi311} \equiv \frac{-ia_2c_2a_3^2}{4}\kappa_{ab}\bar{\Pi}^{\alpha \beta \sigma} k_1^\mu \bar{A}_\alpha{}^a(K_1)e_{\mu \beta}(K_2)\bar{A}_\sigma{}^b(K_3),\\
        &\mathcal{L}_{\phi312} \equiv \frac{-ia_2c_2a_3a_4}{4}\kappa_{ab}\bar{\Pi}^{\alpha \beta \sigma} k_1^\mu \bar{A}_\alpha{}^a(K_1)e_{\mu \beta}(K_2)\bar{f}_\sigma{}^b(K_3), \\
         &\mathcal{L}_{\phi313} \equiv \frac{-ia_2c_2a_3a_4}{4}\kappa_{ab}\bar{\Pi}^{\alpha \beta \sigma} k_1^\mu \bar{f}_\alpha{}^a(K_1)e_{\mu \beta}(K_2)\bar{A}_\sigma{}^b(K_3), \\
        &\mathcal{L}_{\phi314} \equiv \frac{-ia_2c_2a_4^2}{4}\kappa_{ab}\bar{\Pi}^{\alpha \beta \sigma} k_1^\mu \bar{f}_\alpha{}^a(K_1)e_{\mu \beta}(K_2)\bar{f}_\sigma{}^b(K_3),
    \end{aligned}
\end{equation}
with the notation $\bar{f}_\alpha{}^a (K_i) = e_{\nu \alpha}A^{\nu a}(k_i,\bar{k}_i)$. Similarly, starting with $\mathcal{L}_{\phi32}$ and using the double copy maps,
\begin{equation}
    \begin{aligned}
        &\kappa_{AB} \rightarrow a_2 \kappa_{ab},\\
        &f_{abc} \rightarrow d_2 {\Pi}^{\nu \mu \rho}({k}_1,{k}_2,{k}_3),\\
        &\phi^{Aa}(k_i) \rightarrow b_3{A}_\nu{}^a(k_i, \bar{k}_i)+b_4e_{\nu \alpha}\bar{A}^{\alpha b}(k_i,\bar{k}_i),\\
        &\bar{A}_\alpha{}^b(\bar{k}_i) \rightarrow e_{\mu \alpha}(k_i, \bar{k}_i),
    \end{aligned}
\label{cubicdcterms2}
\end{equation}
one gets,
\begin{equation}
    \begin{aligned}
        &\mathcal{L}_{\phi321} \equiv \frac{-ia_2d_2b_3^2}{4}\kappa_{ab}{\Pi}^{\nu \mu \rho} \bar{k}_1^\alpha {A}_\nu{}^a(K_1)e_{\mu \alpha}(K_2){A}_\rho{}^b(K_3),\\
        &\mathcal{L}_{\phi322} \equiv \frac{-ia_2d_2b_3b_4}{4}\kappa_{ab}{\Pi}^{\nu \mu \rho} \bar{k}_1^\alpha {A}_\nu{}^a(K_1)e_{\mu \alpha}(K_2)f_\rho{}^b(K_3), \\
         &\mathcal{L}_{\phi323} \equiv \frac{-ia_2d_2b_3b_4}{4}\kappa_{ab}{\Pi}^{\nu \mu \rho} \bar{k}_1^\alpha f_\nu{}^a(K_1)e_{\mu \alpha}(K_2){A}_\rho{}^b(K_3), \\
        &\mathcal{L}_{\phi324} \equiv \frac{-ia_2d_2b_4^2}{4}\kappa_{ab}{\Pi}^{\nu \mu \rho} \bar{k}_1^\alpha f_\nu{}^a(K_1)e_{\mu \alpha}(K_2)f_\rho{}^b(K_3).
    \end{aligned}
\end{equation}
In (\ref{cubicdcterms2}), the coefficients for the double maps of $\kappa_{AB}$, $\phi^{Aa}$ and $A_\mu{}^b$ are already fixed by the previous results. For the double copy map of $f_{abc}$ we introduce a new coefficient since this map has a different index structure from the one in (\ref{cubicdcterms}). In our construction, we only consider terms up to cubic order, so we calculate the contributions from $\mathcal{L}_{\phi311}$ and $\mathcal{L}_{\phi321}$. The other terms give fourth- and fifth-order terms.
Passing to the position space, $\mathcal{L}_{\phi311}$ gives
\begin{equation}
\begin{aligned}
    \mathcal{B}_6 = &+\frac{ia_2c_2a_3^2}{4}\kappa_{ab}(\bar{\partial}_\sigma e_{\mu \beta})\left((\partial^\mu \bar{A}^{\sigma a})A^{\beta b}-(\partial^\mu \bar{A}^{\beta a})\bar{A}^{\sigma b} \right) \\
    &+\frac{ia_2c_2a_3^2}{4}\kappa_{ab}e_{\mu \beta} (\partial^\mu \bar{A}_\sigma{}^a) \left(\bar{\partial}^\beta \bar{A}^{\sigma b} - \bar{\partial}^\sigma \bar{A}^{\beta b} \right) \\
    &+\frac{ia_2c_2a_3^2}{4}\kappa_{ab}e_{\mu \beta} \bar{A}_\sigma{}^a \left( \bar{\partial}^\sigma \partial^\mu \bar{A}^{\beta b} - \bar{\partial}^\beta \partial^\mu \bar{A}^{\sigma b} \right),
\end{aligned}
\label{b6}
\end{equation}
and $\mathcal{L}_{\phi321}$ gives
\begin{equation}
    \begin{aligned}
        \mathcal{B}_7 = &+\frac{ia_2d_2b_3^2}{4}\kappa_{ab}(\partial_\rho e_{\mu \alpha})\left( (\bar{\partial}^\alpha A^{\rho a})A^{\mu b}-(\bar{\partial}^\alpha A^{\mu a})A^{\rho b} \right) \\
        &+\frac{ia_2d_2b_3^2}{4}\kappa_{ab} e_{\mu \alpha}(\bar{\partial}^\alpha A_\rho{}^a)(\partial^\mu A^{\rho b} - \partial^\rho A^{\mu b}) \\
        &+\frac{ia_2d_2b_3^2}{4}\kappa_{ab}e_{\mu \alpha} A_{\rho}{}^a (\partial^\rho \bar{\partial}^\alpha A^{\mu b} - \partial^\mu \bar{\partial}^\alpha A^{\rho b} ).
    \end{aligned}
    \label{b7}
\end{equation}
After turning of the dual coordinates and using integration by parts, $\mathcal{B}_6$ can be written as
\begin{equation}
    \begin{aligned}
        \mathcal{B}'_6 =\mathcal{C}_1+\mathcal{C}_2+\mathcal{C}_3+\mathcal{C}_4
    \end{aligned}
\end{equation}
where,
\begin{equation}
    \begin{aligned}
        &\mathcal{C}_1 \equiv  -\frac{ia_2c_2a_3^2}{2}\kappa_{ab}(\partial_\sigma e_{\mu \beta})(\partial^\mu A^{\beta a})A^{\sigma b},\\
        &\mathcal{C}_2 \equiv -\frac{ia_2c_2a_3^2}{2}\kappa_{ab} e_{\mu \beta}(\partial^\mu A_{\sigma}{}^a)(\partial^\sigma A^{\beta b}),\\
        &\mathcal{C}_3 \equiv -\frac{ia_2c_2a_3^2}{4}\kappa_{ab} e_{\mu \beta} \left( (\partial^\mu A^{\beta a})(\partial^\sigma A_{\sigma}{}^b)+ (\partial^\mu \partial^\sigma A_\sigma{}^a)A^{\beta b}\right),\\
        &\mathcal{C}_4 \equiv +\frac{ia_2c_2a_3^2}{4}\kappa_{ab}e_{\mu \beta}\left((\partial^\mu A^{\sigma a})(\partial^\beta A_{\sigma}{}^b)-(\partial^\mu \partial^\beta A^{\sigma a})A_{\sigma}{}^b \right)
    \end{aligned}
\label{c1234}
\end{equation}
Similarly, $\mathcal{B}_7$ can be written as
\begin{equation}
    \mathcal{B}'_7 =\mathcal{C}_5+\mathcal{C}_6+\mathcal{C}_7+\mathcal{C}_8,
\end{equation}
where,
\begin{equation}
    \begin{aligned}
        &\mathcal{C}_5 \equiv  -\frac{ia_2d_2b_3^2}{2}\kappa_{ab}(\partial_\sigma e_{\mu \beta})(\partial^\beta A^{\mu a})A^{\sigma b},\\
        &\mathcal{C}_6 \equiv -\frac{ia_2d_2b_3^2}{2}\kappa_{ab} e_{\mu \beta}(\partial^\mu A_{\sigma}{}^a)(\partial^\sigma A^{\beta b}),\\
        &\mathcal{C}_7 \equiv -\frac{ia_2d_2b_3^2}{4}\kappa_{ab} e_{\mu \beta} \left( (\partial^\beta A^{\mu a})(\partial^\sigma A_{\sigma}{}^b)+ (\partial^\beta \partial^\sigma A_\sigma{}^a)A^{\mu b}\right),\\
        &\mathcal{C}_8 \equiv +\frac{ia_2d_2b_3^2}{4}\kappa_{ab}e_{\mu \beta}\left((\partial^\mu A^{\sigma a})(\partial^\beta A_{\sigma}{}^b)-(\partial^\mu \partial^\beta A^{\sigma a})A_{\sigma}{}^b \right)
    \end{aligned}
    \label{c5678}
\end{equation}
Now, we can compare the terms in (\ref{c1234}) and (\ref{c5678}) with the terms in (\ref{heterodftlimitaction1}) to impose further constraints on the coefficients. Choosing the constants in such a way that
\begin{equation}
    a_2c_2a_3^2=\frac{i}{2} \quad \text{and} \quad a_2d_2b_3^2=-\frac{i}{2},
\label{thirdconstraint}
\end{equation}
$\mathcal{C}_1$ gives the fifth term of (\ref{eAAterms}) and $\mathcal{C}_5$ gives the fourth term of (\ref{eAAterms}). Note that these were the only missing terms of (\ref{eAAterms}) in our double copy construction. Hence until now, we constructed all the terms (\ref{usualdftterms}), (\ref{eAAterms}) and (\ref{ymterms}). \\

Before commenting on the double copy construction of the terms (\ref{AAAAterms}) and (\ref{higherorderterms}), we should also analyze the extra terms coming from our construction. These extra terms are given by $\mathcal{B}_5$ in (\ref{b5extra}), $\mathcal{C}_2, \mathcal{C}_3, \mathcal{C}_4 $ from (\ref{c1234}), and $\mathcal{C}_6,\mathcal{C}_7,\mathcal{C}_8$ from (\ref{c5678}). When we impose the constraint (\ref{thirdconstraint}), $\mathcal{C}_2$ and $\mathcal{C}_6$ exactly cancel the extra term $\mathcal{B}_5$. Similarly, $\mathcal{C}_4$ and $\mathcal{C}_8$ cancel each other. So, we only remain with the extra terms $\mathcal{C}_3, \mathcal{C}_7$. If we impose Lorenz gauge, namely,
\begin{equation}
    \partial^\mu A_\mu = 0,
\label{lorenzgauge}
\end{equation}
$\mathcal{C}_3$ and $\mathcal{C}_7$ also vanish. \\
Let us summarize what we have done so far. Using double copy construction on the terms $\mathcal{L}_{\phi 2}$, $\mathcal{L}_{\phi 3}$, $\mathcal{L}_{\phi 4}$ from (\ref{fivelag}); we constructed the terms (\ref{usualdftterms}), (\ref{eAAterms}) and (\ref{ymterms}) from (\ref{heterodftlimitaction1}). There remains no extra term when we impose Lorenz gauge (\ref{lorenzgauge}). In our construction, we use arbitrary coefficients in the double copy maps, but obtained some constraints in these coefficients. The arbitrary coefficients we introduced are given as $a_1, a_2, a_3, a_4, b_1, b_3, b_4, c_2$ and $d_2$, with constraints,
\begin{equation}
    \begin{aligned}
        &a_1a_2a_3^2+b_1a_2b_3^2 = -2, \\
        &a_1a_2a_3a_4 = -\frac{1}{2} \quad \text{and} \quad b_1a_2b_3b_4=+\frac{1}{2}, \\
        &a_2c_2a_3^2=\frac{i}{2} \quad \text{and} \quad a_2d_2b_3^2=-\frac{i}{2}.
    \end{aligned}
\label{allconstraints}
\end{equation}
Obviously, one can introduce a set of numbers to these coefficients, in a way that these constraints are satisfied. Further constraints can be obtained from an analysis of higher-order terms and extension of the construction to the non-Abelian sector. For now, let us choose a set for these coefficients to discuss their implications for the double copy procedure. For instance, let us choose the coefficients as,
\begin{equation}
    \begin{aligned}
       & a_3 = 1, \quad a_4 = \frac{1}{2}, \quad b_3 = 1, \quad b_4 =- \frac{1}{2}, \quad a_2 = -2,\\
       & a_1 = \frac{1}{2}, \quad b_1 = \frac{1}{2}, \quad c_2 =-\frac{i}{4}, \quad d_2 = \frac{i}{4}.
    \end{aligned}
\end{equation}
The different choices for $c_2$ and $d_2$ imply two different maps as
\begin{equation}
    f_{abc} \rightarrow -\frac{i}{4} \bar{\Pi}^{\alpha \beta \sigma}, \quad \text{and} \quad  f_{abc} \rightarrow +\frac{i}{4} {\Pi}^{\mu \nu \rho}.
\end{equation}
Since these two maps have different index structures, one with barred indices and other with unbarred indices, we believe that it is possible that they can have different coefficients. However, this situation should be understood better and as we have already mentioned further analysis of higher order terms are necessary to impose more constraints on the coefficients. So, it is a good point to discuss the higher order terms. \\

Consider $\mathcal{A}_3$ terms given in (\ref{AAAAterms}). Looking at their structures, one can see that these terms may come from the $\mathcal{L}_{\phi5}$ piece from (\ref{fivelag}). The problem in this piece is the existence of an inverse of Cartan-Killing metric, namely $\kappa^{ef}$. As it was also discussed in \cite{jaram}, the double copy map (\ref{kappadc}) introduced for the Cartan-Killing metric is not invertible. In that sense it is not obvious how to define a double copy map for $\kappa^{ef}$. One possibility is to start the construction by choosing a gauge in a way that the propagator of the theory becomes invertible. However, one can still continue with the structural form of the double copy of $\mathcal{L}_{\phi5}$ to check if it gives consistent terms. After inserting the double copy maps (\ref{dcmap}), (\ref{fdcmap}), (\ref{ansatzdc}) into $\mathcal{L}_{\phi 5}$, the first term takes the structural form
\begin{equation}
    \mathcal{L}_{\phi 5} \sim \kappa_{ac} \,\kappa_{bd} \, \eta_{mn} \bar{\Pi}^{\alpha \beta m} \bar{\Pi}^{n\sigma \tau} A_\alpha{}^aA_\beta{}^bA_\sigma{}^cA_\tau{}^d.
\end{equation}
This construction is incomplete since each $\bar{\Pi}^{klm}$ depends on three momenta, while in the above term there are only four $A$ fields. However, one can see that such a construction creates many of the $\mathcal{A}_3$ terms given in (\ref{AAAAterms}). \\

To sum up, to fix the coefficients by imposing more constraints one needs an analysis of the higher-order terms, which is also problematic due to know limitations of double copy construction of DFT.
\section{Discussion and Conclusion}
In this work, a double copy procedure for the Abelian subsector of DFT action is offered. Inspired by the results of \cite{expform}, the starting point is taken as the ($YM + \phi^3 $) theory introduced in \cite{expform}. In that work \cite{expform}, the authors show that the scattering amplitudes for the YME theory coupled to the dilaton and two-form field can be obtained as a double copy of pure Yang-Mils theory and ($YM + \phi^3 $). Since the usual DFT action is reduced to the supergravity action when the dependence on dual coordinates are turned off, it is consistent to use this double copy idea for the heterotic DFT. \\

To obtain the Abelian sector of the heterotic DFT formulation, we choose the structure constants of the usual gauge fields, $f_{abc}$, nonzero while choosing $F_{ABC}=0$. The motivation of this choice is based on the idea that the color indices $a,b,c...$ are mapped to the dual indices of the DFT fields, so their presence does not make the final theory non-Abelian. On the other hand, in this method $A,B,C...$ indices are mapped to the $a,b,c...$ color indices, so choosing $F_{ABC}=0$ makes the final double copy theory Abelian. Using this formalism, together with the double copy maps (\ref{dcmap}),(\ref{fdcmap}), (\ref{kappadc}), (\ref{ansatzdc}), (\ref{quadsymdcmaps}), (\ref{cubicdcterms}), (\ref{cubicdcterms2}) all the terms (up to cubic order) in the Abelian sector of heterotic DFT action are obtained. The double copy maps are introduced with arbitrary coefficients at the beginning, but at the end some constraints (\ref{allconstraints}) are obtained for an exact match. Further constraints on these coefficients should come from the analysis of higher order terms, but this analysis is problematic because of the known limitations of double copy construction of DFT. \\

These limitations for the higher order analysis can be seen in the last two terms in (\ref{fivelag}), namely $\mathcal{L}_{\phi 4}$ and $\mathcal{L}_{\phi5}$. While $\mathcal{L}_{\phi 4}$ gives structurally $(\partial e)A(\partial e)A$ terms, $\mathcal{L}_{\phi 5}$ is expected to give $(\partial A)(\partial A)AA$ terms. The problem with these terms is the presence of an inverse Cartan-Killing metric, namely $\kappa^{ab}$. The double copy map for $\kappa_{ab}$ is given in (\ref{dcmap}) and one can see that it is not invertible. In that sense, it is not obvious how to define a double copy map for the inverse Cartan-Killing metric, $\kappa^{ab}$. Moreover, the structural forms of the double copy of these two terms seem to depend on six momenta, while the field contents include only four fields. So, the double copy procedure for the last two pieces of (\ref{fivelag}) is incomplete. However, one can still see that even this problematic structure of the double copy of $\mathcal{L}_{\phi5}$ gives many of the terms in (\ref{AAAAterms}). A possible solution for this problematic structure can be to bring the action to the cubic form by introducing some auxiliary fields, as it was suggested in \cite{Bern2010}.\\

To summarize, in this work we offer a double copy map for the Abelian sector of heterotic DFT starting with a pure Yang-Mills theory and $YM+\phi^3$ theory, inspired by the results of \cite{expform}, \cite{Lescano:2024gma}. The action we constructed gives an exact match with the Abelian sector of heterotic DFT up to cubic order. In that sense the results are promising, but it is also important to understand the origin of the double copy maps we introduced. Some strategies we use when introducing the maps are unusual to the known literature, for example introducing two different maps for $f_{abc}$ for barred and unbarred indices. However, obtaining an exact match with no extra terms implies that our strategy can be useful for future works. Extending these results to the non-Abelian sector and to the higher order terms needs more careful analysis, which we postpone to the later works. \\

\begin{acknowledgments}
I would like to express my deepest gratitude to Özgür Sarıoğlu for the helpful discussions, and continuous support he gave throughout the development of this article. I am also grateful to Eric Lescano and Jesus Rodriguez for their comments and valuable suggestions, and Kanghoon Lee for pointing out some important references.
\end{acknowledgments}

\bibliography{apssamp}% Produces the bibliography via BibTeX.

\begin{thebibliography}{37}%
\makeatletter
\providecommand \@ifxundefined [1]{%
 \@ifx{#1\undefined}
}%
\providecommand \@ifnum [1]{%
 \ifnum #1\expandafter \@firstoftwo
 \else \expandafter \@secondoftwo
 \fi
}%
\providecommand \@ifx [1]{%
 \ifx #1\expandafter \@firstoftwo
 \else \expandafter \@secondoftwo
 \fi
}%
\providecommand \natexlab [1]{#1}%
\providecommand \enquote  [1]{``#1''}%
\providecommand \bibnamefont  [1]{#1}%
\providecommand \bibfnamefont [1]{#1}%
\providecommand \citenamefont [1]{#1}%
\providecommand \href@noop [0]{\@secondoftwo}%
\providecommand \href [0]{\begingroup \@sanitize@url \@href}%
\providecommand \@href[1]{\@@startlink{#1}\@@href}%
\providecommand \@@href[1]{\endgroup#1\@@endlink}%
\providecommand \@sanitize@url [0]{\catcode `\\12\catcode `\$12\catcode `\&12\catcode `\#12\catcode `\^12\catcode `\_12\catcode `\%12\relax}%
\providecommand \@@startlink[1]{}%
\providecommand \@@endlink[0]{}%
\providecommand \url  [0]{\begingroup\@sanitize@url \@url }%
\providecommand \@url [1]{\endgroup\@href {#1}{\urlprefix }}%
\providecommand \urlprefix  [0]{URL }%
\providecommand \Eprint [0]{\href }%
\providecommand \doibase [0]{https://doi.org/}%
\providecommand \selectlanguage [0]{\@gobble}%
\providecommand \bibinfo  [0]{\@secondoftwo}%
\providecommand \bibfield  [0]{\@secondoftwo}%
\providecommand \translation [1]{[#1]}%
\providecommand \BibitemOpen [0]{}%
\providecommand \bibitemStop [0]{}%
\providecommand \bibitemNoStop [0]{.\EOS\space}%
\providecommand \EOS [0]{\spacefactor3000\relax}%
\providecommand \BibitemShut  [1]{\csname bibitem#1\endcsname}%
\let\auto@bib@innerbib\@empty
%</preamble>
\bibitem [{\citenamefont {Bern}\ \emph {et~al.}(2008)\citenamefont {Bern}, \citenamefont {Carrasco},\ and\ \citenamefont {Johansson}}]{bcj}%
  \BibitemOpen
  \bibfield  {author} {\bibinfo {author} {\bibfnamefont {Z.}~\bibnamefont {Bern}}, \bibinfo {author} {\bibfnamefont {J.~J.~M.}\ \bibnamefont {Carrasco}},\ and\ \bibinfo {author} {\bibfnamefont {H.}~\bibnamefont {Johansson}},\ }\bibfield  {title} {\bibinfo {title} {{New Relations for Gauge-Theory Amplitudes}},\ }\href {https://doi.org/10.1103/PhysRevD.78.085011} {\bibfield  {journal} {\bibinfo  {journal} {Phys. Rev. D}\ }\textbf {\bibinfo {volume} {78}},\ \bibinfo {pages} {085011} (\bibinfo {year} {2008})},\ \Eprint {https://arxiv.org/abs/0805.3993} {arXiv:0805.3993 [hep-ph]} \BibitemShut {NoStop}%
\bibitem [{\citenamefont {Bern}\ \emph {et~al.}(2010{\natexlab{a}})\citenamefont {Bern}, \citenamefont {Carrasco},\ and\ \citenamefont {Johansson}}]{bcj2}%
  \BibitemOpen
  \bibfield  {author} {\bibinfo {author} {\bibfnamefont {Z.}~\bibnamefont {Bern}}, \bibinfo {author} {\bibfnamefont {J.~J.~M.}\ \bibnamefont {Carrasco}},\ and\ \bibinfo {author} {\bibfnamefont {H.}~\bibnamefont {Johansson}},\ }\bibfield  {title} {\bibinfo {title} {{Perturbative Quantum Gravity as a Double Copy of Gauge Theory}},\ }\href {https://doi.org/10.1103/PhysRevLett.105.061602} {\bibfield  {journal} {\bibinfo  {journal} {Phys. Rev. Lett.}\ }\textbf {\bibinfo {volume} {105}},\ \bibinfo {pages} {061602} (\bibinfo {year} {2010}{\natexlab{a}})},\ \Eprint {https://arxiv.org/abs/1004.0476} {arXiv:1004.0476 [hep-th]} \BibitemShut {NoStop}%
\bibitem [{\citenamefont {Kawai}\ \emph {et~al.}(1986)\citenamefont {Kawai}, \citenamefont {Lewellen},\ and\ \citenamefont {Tye}}]{klt}%
  \BibitemOpen
  \bibfield  {author} {\bibinfo {author} {\bibfnamefont {H.}~\bibnamefont {Kawai}}, \bibinfo {author} {\bibfnamefont {D.~C.}\ \bibnamefont {Lewellen}},\ and\ \bibinfo {author} {\bibfnamefont {S.~H.~H.}\ \bibnamefont {Tye}},\ }\bibfield  {title} {\bibinfo {title} {{A Relation Between Tree Amplitudes of Closed and Open Strings}},\ }\href {https://doi.org/10.1016/0550-3213(86)90362-7} {\bibfield  {journal} {\bibinfo  {journal} {Nucl. Phys. B}\ }\textbf {\bibinfo {volume} {269}},\ \bibinfo {pages} {1} (\bibinfo {year} {1986})}\BibitemShut {NoStop}%
\bibitem [{\citenamefont {Bern}\ \emph {et~al.}(2010{\natexlab{b}})\citenamefont {Bern}, \citenamefont {Dennen}, \citenamefont {Huang},\ and\ \citenamefont {Kiermaier}}]{Bern2010}%
  \BibitemOpen
  \bibfield  {author} {\bibinfo {author} {\bibfnamefont {Z.}~\bibnamefont {Bern}}, \bibinfo {author} {\bibfnamefont {T.}~\bibnamefont {Dennen}}, \bibinfo {author} {\bibfnamefont {Y.-t.}\ \bibnamefont {Huang}},\ and\ \bibinfo {author} {\bibfnamefont {M.}~\bibnamefont {Kiermaier}},\ }\bibfield  {title} {\bibinfo {title} {{Gravity as the Square of Gauge Theory}},\ }\href {https://doi.org/10.1103/PhysRevD.82.065003} {\bibfield  {journal} {\bibinfo  {journal} {Phys. Rev. D}\ }\textbf {\bibinfo {volume} {82}},\ \bibinfo {pages} {065003} (\bibinfo {year} {2010}{\natexlab{b}})},\ \Eprint {https://arxiv.org/abs/1004.0693} {arXiv:1004.0693 [hep-th]} \BibitemShut {NoStop}%
\bibitem [{\citenamefont {Johansson}\ and\ \citenamefont {Nohle}(2017)}]{johan}%
  \BibitemOpen
  \bibfield  {author} {\bibinfo {author} {\bibfnamefont {H.}~\bibnamefont {Johansson}}\ and\ \bibinfo {author} {\bibfnamefont {J.}~\bibnamefont {Nohle}},\ }\bibfield  {title} {\bibinfo {title} {{Conformal Gravity from Gauge Theory}},\ }\href@noop {} {\  (\bibinfo {year} {2017})},\ \Eprint {https://arxiv.org/abs/1707.02965} {arXiv:1707.02965 [hep-th]} \BibitemShut {NoStop}%
\bibitem [{\citenamefont {Johansson}\ \emph {et~al.}(2018)\citenamefont {Johansson}, \citenamefont {Mogull},\ and\ \citenamefont {Teng}}]{johan2}%
  \BibitemOpen
  \bibfield  {author} {\bibinfo {author} {\bibfnamefont {H.}~\bibnamefont {Johansson}}, \bibinfo {author} {\bibfnamefont {G.}~\bibnamefont {Mogull}},\ and\ \bibinfo {author} {\bibfnamefont {F.}~\bibnamefont {Teng}},\ }\bibfield  {title} {\bibinfo {title} {{Unraveling conformal gravity amplitudes}},\ }\href {https://doi.org/10.1007/JHEP09(2018)080} {\bibfield  {journal} {\bibinfo  {journal} {JHEP}\ }\textbf {\bibinfo {volume} {09}},\ \bibinfo {pages} {080}},\ \Eprint {https://arxiv.org/abs/1806.05124} {arXiv:1806.05124 [hep-th]} \BibitemShut {NoStop}%
\bibitem [{\citenamefont {Luna}\ \emph {et~al.}(2015)\citenamefont {Luna}, \citenamefont {Monteiro}, \citenamefont {O'Connell},\ and\ \citenamefont {White}}]{Luna}%
  \BibitemOpen
  \bibfield  {author} {\bibinfo {author} {\bibfnamefont {A.}~\bibnamefont {Luna}}, \bibinfo {author} {\bibfnamefont {R.}~\bibnamefont {Monteiro}}, \bibinfo {author} {\bibfnamefont {D.}~\bibnamefont {O'Connell}},\ and\ \bibinfo {author} {\bibfnamefont {C.~D.}\ \bibnamefont {White}},\ }\bibfield  {title} {\bibinfo {title} {{The classical double copy for Taub\textendash{}NUT spacetime}},\ }\href {https://doi.org/10.1016/j.physletb.2015.09.021} {\bibfield  {journal} {\bibinfo  {journal} {Phys. Lett. B}\ }\textbf {\bibinfo {volume} {750}},\ \bibinfo {pages} {272} (\bibinfo {year} {2015})},\ \Eprint {https://arxiv.org/abs/1507.01869} {arXiv:1507.01869 [hep-th]} \BibitemShut {NoStop}%
\bibitem [{\citenamefont {Carrillo-Gonz\'alez}\ \emph {et~al.}(2018)\citenamefont {Carrillo-Gonz\'alez}, \citenamefont {Penco},\ and\ \citenamefont {Trodden}}]{Carrillo}%
  \BibitemOpen
  \bibfield  {author} {\bibinfo {author} {\bibfnamefont {M.}~\bibnamefont {Carrillo-Gonz\'alez}}, \bibinfo {author} {\bibfnamefont {R.}~\bibnamefont {Penco}},\ and\ \bibinfo {author} {\bibfnamefont {M.}~\bibnamefont {Trodden}},\ }\bibfield  {title} {\bibinfo {title} {{The classical double copy in maximally symmetric spacetimes}},\ }\href {https://doi.org/10.1007/JHEP04(2018)028} {\bibfield  {journal} {\bibinfo  {journal} {JHEP}\ }\textbf {\bibinfo {volume} {04}},\ \bibinfo {pages} {028}},\ \Eprint {https://arxiv.org/abs/1711.01296} {arXiv:1711.01296 [hep-th]} \BibitemShut {NoStop}%
\bibitem [{\citenamefont {Bahjat-Abbas}\ \emph {et~al.}(2017)\citenamefont {Bahjat-Abbas}, \citenamefont {Luna},\ and\ \citenamefont {White}}]{abbas}%
  \BibitemOpen
  \bibfield  {author} {\bibinfo {author} {\bibfnamefont {N.}~\bibnamefont {Bahjat-Abbas}}, \bibinfo {author} {\bibfnamefont {A.}~\bibnamefont {Luna}},\ and\ \bibinfo {author} {\bibfnamefont {C.~D.}\ \bibnamefont {White}},\ }\bibfield  {title} {\bibinfo {title} {{The Kerr-Schild double copy in curved spacetime}},\ }\href {https://doi.org/10.1007/JHEP12(2017)004} {\bibfield  {journal} {\bibinfo  {journal} {JHEP}\ }\textbf {\bibinfo {volume} {12}},\ \bibinfo {pages} {004}},\ \Eprint {https://arxiv.org/abs/1710.01953} {arXiv:1710.01953 [hep-th]} \BibitemShut {NoStop}%
\bibitem [{\citenamefont {Adamo}\ \emph {et~al.}(2022)\citenamefont {Adamo}, \citenamefont {Carrasco}, \citenamefont {Carrillo-Gonz\'alez}, \citenamefont {Chiodaroli}, \citenamefont {Elvang}, \citenamefont {Johansson}, \citenamefont {O'Connell}, \citenamefont {Roiban},\ and\ \citenamefont {Schlotterer}}]{adamo}%
  \BibitemOpen
  \bibfield  {author} {\bibinfo {author} {\bibfnamefont {T.}~\bibnamefont {Adamo}}, \bibinfo {author} {\bibfnamefont {J.~J.~M.}\ \bibnamefont {Carrasco}}, \bibinfo {author} {\bibfnamefont {M.}~\bibnamefont {Carrillo-Gonz\'alez}}, \bibinfo {author} {\bibfnamefont {M.}~\bibnamefont {Chiodaroli}}, \bibinfo {author} {\bibfnamefont {H.}~\bibnamefont {Elvang}}, \bibinfo {author} {\bibfnamefont {H.}~\bibnamefont {Johansson}}, \bibinfo {author} {\bibfnamefont {D.}~\bibnamefont {O'Connell}}, \bibinfo {author} {\bibfnamefont {R.}~\bibnamefont {Roiban}},\ and\ \bibinfo {author} {\bibfnamefont {O.}~\bibnamefont {Schlotterer}},\ }\bibfield  {title} {\bibinfo {title} {{Snowmass White Paper: the Double Copy and its Applications}},\ }in\ \href@noop {} {\emph {\bibinfo {booktitle} {{Snowmass 2021}}}}\ (\bibinfo {year} {2022})\ \Eprint {https://arxiv.org/abs/2204.06547} {arXiv:2204.06547 [hep-th]} \BibitemShut {NoStop}%
\bibitem [{\citenamefont {Chiodaroli}\ \emph {et~al.}(2017)\citenamefont {Chiodaroli}, \citenamefont {Gunaydin}, \citenamefont {Johansson},\ and\ \citenamefont {Roiban}}]{expform}%
  \BibitemOpen
  \bibfield  {author} {\bibinfo {author} {\bibfnamefont {M.}~\bibnamefont {Chiodaroli}}, \bibinfo {author} {\bibfnamefont {M.}~\bibnamefont {Gunaydin}}, \bibinfo {author} {\bibfnamefont {H.}~\bibnamefont {Johansson}},\ and\ \bibinfo {author} {\bibfnamefont {R.}~\bibnamefont {Roiban}},\ }\bibfield  {title} {\bibinfo {title} {{Explicit Formulae for Yang-Mills-Einstein Amplitudes from the Double Copy}},\ }\href {https://doi.org/10.1007/JHEP07(2017)002} {\bibfield  {journal} {\bibinfo  {journal} {JHEP}\ }\textbf {\bibinfo {volume} {07}},\ \bibinfo {pages} {002}},\ \Eprint {https://arxiv.org/abs/1703.00421} {arXiv:1703.00421 [hep-th]} \BibitemShut {NoStop}%
\bibitem [{\citenamefont {Lee}(2018{\natexlab{a}})}]{Leekerr}%
  \BibitemOpen
  \bibfield  {author} {\bibinfo {author} {\bibfnamefont {K.}~\bibnamefont {Lee}},\ }\bibfield  {title} {\bibinfo {title} {{Kerr-Schild Double Field Theory and Classical Double Copy}},\ }\href {https://doi.org/10.1007/JHEP10(2018)027} {\bibfield  {journal} {\bibinfo  {journal} {JHEP}\ }\textbf {\bibinfo {volume} {10}},\ \bibinfo {pages} {027}},\ \Eprint {https://arxiv.org/abs/1807.08443} {arXiv:1807.08443 [hep-th]} \BibitemShut {NoStop}%
\bibitem [{\citenamefont {Cho}\ \emph {et~al.}(2022)\citenamefont {Cho}, \citenamefont {Kim},\ and\ \citenamefont {Lee}}]{Cholee}%
  \BibitemOpen
  \bibfield  {author} {\bibinfo {author} {\bibfnamefont {K.}~\bibnamefont {Cho}}, \bibinfo {author} {\bibfnamefont {K.}~\bibnamefont {Kim}},\ and\ \bibinfo {author} {\bibfnamefont {K.}~\bibnamefont {Lee}},\ }\bibfield  {title} {\bibinfo {title} {{The off-shell recursion for gravity and the classical double copy for currents}},\ }\href {https://doi.org/10.1007/JHEP01(2022)186} {\bibfield  {journal} {\bibinfo  {journal} {JHEP}\ }\textbf {\bibinfo {volume} {01}},\ \bibinfo {pages} {186}},\ \Eprint {https://arxiv.org/abs/2109.06392} {arXiv:2109.06392 [hep-th]} \BibitemShut {NoStop}%
\bibitem [{\citenamefont {Kim}\ \emph {et~al.}(2020)\citenamefont {Kim}, \citenamefont {Lee}, \citenamefont {Monteiro}, \citenamefont {Nicholson},\ and\ \citenamefont {Peinador~Veiga}}]{Kim}%
  \BibitemOpen
  \bibfield  {author} {\bibinfo {author} {\bibfnamefont {K.}~\bibnamefont {Kim}}, \bibinfo {author} {\bibfnamefont {K.}~\bibnamefont {Lee}}, \bibinfo {author} {\bibfnamefont {R.}~\bibnamefont {Monteiro}}, \bibinfo {author} {\bibfnamefont {I.}~\bibnamefont {Nicholson}},\ and\ \bibinfo {author} {\bibfnamefont {D.}~\bibnamefont {Peinador~Veiga}},\ }\bibfield  {title} {\bibinfo {title} {{The Classical Double Copy of a Point Charge}},\ }\href {https://doi.org/10.1007/JHEP02(2020)046} {\bibfield  {journal} {\bibinfo  {journal} {JHEP}\ }\textbf {\bibinfo {volume} {02}},\ \bibinfo {pages} {046}},\ \Eprint {https://arxiv.org/abs/1912.02177} {arXiv:1912.02177 [hep-th]} \BibitemShut {NoStop}%
\bibitem [{\citenamefont {Jonke}\ and\ \citenamefont {Lescano}(2025)}]{lescanononcom}%
  \BibitemOpen
  \bibfield  {author} {\bibinfo {author} {\bibfnamefont {L.}~\bibnamefont {Jonke}}\ and\ \bibinfo {author} {\bibfnamefont {E.}~\bibnamefont {Lescano}},\ }\bibfield  {title} {\bibinfo {title} {{From noncommutative Yang-Mills to noncommutative gravity through a classical double copy map}},\ }\href@noop {} {\  (\bibinfo {year} {2025})},\ \Eprint {https://arxiv.org/abs/2502.03521} {arXiv:2502.03521 [hep-th]} \BibitemShut {NoStop}%
\bibitem [{\citenamefont {Diaz-Jaramillo}\ \emph {et~al.}(2022)\citenamefont {Diaz-Jaramillo}, \citenamefont {Hohm},\ and\ \citenamefont {Plefka}}]{jaram}%
  \BibitemOpen
  \bibfield  {author} {\bibinfo {author} {\bibfnamefont {F.}~\bibnamefont {Diaz-Jaramillo}}, \bibinfo {author} {\bibfnamefont {O.}~\bibnamefont {Hohm}},\ and\ \bibinfo {author} {\bibfnamefont {J.}~\bibnamefont {Plefka}},\ }\bibfield  {title} {\bibinfo {title} {{Double field theory as the double copy of Yang-Mills theory}},\ }\href {https://doi.org/10.1103/PhysRevD.105.045012} {\bibfield  {journal} {\bibinfo  {journal} {Phys. Rev. D}\ }\textbf {\bibinfo {volume} {105}},\ \bibinfo {pages} {045012} (\bibinfo {year} {2022})},\ \Eprint {https://arxiv.org/abs/2109.01153} {arXiv:2109.01153 [hep-th]} \BibitemShut {NoStop}%
\bibitem [{\citenamefont {Ferrero}\ and\ \citenamefont {Francia}(2021)}]{dario}%
  \BibitemOpen
  \bibfield  {author} {\bibinfo {author} {\bibfnamefont {P.}~\bibnamefont {Ferrero}}\ and\ \bibinfo {author} {\bibfnamefont {D.}~\bibnamefont {Francia}},\ }\bibfield  {title} {\bibinfo {title} {{On the Lagrangian formulation of the double copy to cubic order}},\ }\href {https://doi.org/10.1007/JHEP02(2021)213} {\bibfield  {journal} {\bibinfo  {journal} {JHEP}\ }\textbf {\bibinfo {volume} {02}},\ \bibinfo {pages} {213}},\ \Eprint {https://arxiv.org/abs/2012.00713} {arXiv:2012.00713 [hep-th]} \BibitemShut {NoStop}%
\bibitem [{\citenamefont {Hull}\ and\ \citenamefont {Zwiebach}(2009)}]{hulldft}%
  \BibitemOpen
  \bibfield  {author} {\bibinfo {author} {\bibfnamefont {C.}~\bibnamefont {Hull}}\ and\ \bibinfo {author} {\bibfnamefont {B.}~\bibnamefont {Zwiebach}},\ }\bibfield  {title} {\bibinfo {title} {{Double Field Theory}},\ }\href {https://doi.org/10.1088/1126-6708/2009/09/099} {\bibfield  {journal} {\bibinfo  {journal} {JHEP}\ }\textbf {\bibinfo {volume} {09}},\ \bibinfo {pages} {099}},\ \Eprint {https://arxiv.org/abs/0904.4664} {arXiv:0904.4664 [hep-th]} \BibitemShut {NoStop}%
\bibitem [{\citenamefont {Hohm}\ \emph {et~al.}(2010{\natexlab{a}})\citenamefont {Hohm}, \citenamefont {Hull},\ and\ \citenamefont {Zwiebach}}]{backindepdft}%
  \BibitemOpen
  \bibfield  {author} {\bibinfo {author} {\bibfnamefont {O.}~\bibnamefont {Hohm}}, \bibinfo {author} {\bibfnamefont {C.}~\bibnamefont {Hull}},\ and\ \bibinfo {author} {\bibfnamefont {B.}~\bibnamefont {Zwiebach}},\ }\bibfield  {title} {\bibinfo {title} {{Background independent action for double field theory}},\ }\href {https://doi.org/10.1007/JHEP07(2010)016} {\bibfield  {journal} {\bibinfo  {journal} {JHEP}\ }\textbf {\bibinfo {volume} {07}},\ \bibinfo {pages} {016}},\ \Eprint {https://arxiv.org/abs/1003.5027} {arXiv:1003.5027 [hep-th]} \BibitemShut {NoStop}%
\bibitem [{\citenamefont {Hohm}\ \emph {et~al.}(2010{\natexlab{b}})\citenamefont {Hohm}, \citenamefont {Hull},\ and\ \citenamefont {Zwiebach}}]{hohmdft}%
  \BibitemOpen
  \bibfield  {author} {\bibinfo {author} {\bibfnamefont {O.}~\bibnamefont {Hohm}}, \bibinfo {author} {\bibfnamefont {C.}~\bibnamefont {Hull}},\ and\ \bibinfo {author} {\bibfnamefont {B.}~\bibnamefont {Zwiebach}},\ }\bibfield  {title} {\bibinfo {title} {{Generalized metric formulation of double field theory}},\ }\href {https://doi.org/10.1007/JHEP08(2010)008} {\bibfield  {journal} {\bibinfo  {journal} {JHEP}\ }\textbf {\bibinfo {volume} {08}},\ \bibinfo {pages} {008}},\ \Eprint {https://arxiv.org/abs/1006.4823} {arXiv:1006.4823 [hep-th]} \BibitemShut {NoStop}%
\bibitem [{\citenamefont {Aldazabal}\ \emph {et~al.}(2013)\citenamefont {Aldazabal}, \citenamefont {Marques},\ and\ \citenamefont {Nunez}}]{aldaz}%
  \BibitemOpen
  \bibfield  {author} {\bibinfo {author} {\bibfnamefont {G.}~\bibnamefont {Aldazabal}}, \bibinfo {author} {\bibfnamefont {D.}~\bibnamefont {Marques}},\ and\ \bibinfo {author} {\bibfnamefont {C.}~\bibnamefont {Nunez}},\ }\bibfield  {title} {\bibinfo {title} {{Double Field Theory: A Pedagogical Review}},\ }\href {https://doi.org/10.1088/0264-9381/30/16/163001} {\bibfield  {journal} {\bibinfo  {journal} {Class. Quant. Grav.}\ }\textbf {\bibinfo {volume} {30}},\ \bibinfo {pages} {163001} (\bibinfo {year} {2013})},\ \Eprint {https://arxiv.org/abs/1305.1907} {arXiv:1305.1907 [hep-th]} \BibitemShut {NoStop}%
\bibitem [{\citenamefont {Lee}(2018{\natexlab{b}})}]{Lee}%
  \BibitemOpen
  \bibfield  {author} {\bibinfo {author} {\bibfnamefont {K.}~\bibnamefont {Lee}},\ }\bibfield  {title} {\bibinfo {title} {{Kerr-Schild Double Field Theory and Classical Double Copy}},\ }\href {https://doi.org/10.1007/JHEP10(2018)027} {\bibfield  {journal} {\bibinfo  {journal} {JHEP}\ }\textbf {\bibinfo {volume} {10}},\ \bibinfo {pages} {027}},\ \Eprint {https://arxiv.org/abs/1807.08443} {arXiv:1807.08443 [hep-th]} \BibitemShut {NoStop}%
\bibitem [{\citenamefont {Cho}\ and\ \citenamefont {Lee}(2019{\natexlab{a}})}]{Cho}%
  \BibitemOpen
  \bibfield  {author} {\bibinfo {author} {\bibfnamefont {W.}~\bibnamefont {Cho}}\ and\ \bibinfo {author} {\bibfnamefont {K.}~\bibnamefont {Lee}},\ }\bibfield  {title} {\bibinfo {title} {{Heterotic Kerr-Schild Double Field Theory and Classical Double Copy}},\ }\href {https://doi.org/10.1007/JHEP07(2019)030} {\bibfield  {journal} {\bibinfo  {journal} {JHEP}\ }\textbf {\bibinfo {volume} {07}},\ \bibinfo {pages} {030}},\ \Eprint {https://arxiv.org/abs/1904.11650} {arXiv:1904.11650 [hep-th]} \BibitemShut {NoStop}%
\bibitem [{\citenamefont {Lescano}\ and\ \citenamefont {Roychowdhury}(2022)}]{lescanohetero}%
  \BibitemOpen
  \bibfield  {author} {\bibinfo {author} {\bibfnamefont {E.}~\bibnamefont {Lescano}}\ and\ \bibinfo {author} {\bibfnamefont {S.}~\bibnamefont {Roychowdhury}},\ }\bibfield  {title} {\bibinfo {title} {{Heterotic Kerr-Schild Double Field Theory and its double Yang-Mills formulation}},\ }\href {https://doi.org/10.1007/JHEP04(2022)090} {\bibfield  {journal} {\bibinfo  {journal} {JHEP}\ }\textbf {\bibinfo {volume} {04}},\ \bibinfo {pages} {090}},\ \Eprint {https://arxiv.org/abs/2201.09364} {arXiv:2201.09364 [hep-th]} \BibitemShut {NoStop}%
\bibitem [{\citenamefont {Lescano}\ \emph {et~al.}(2023)\citenamefont {Lescano}, \citenamefont {Menezes},\ and\ \citenamefont {Rodr\'\i{}guez}}]{Lescano:2023pai}%
  \BibitemOpen
  \bibfield  {author} {\bibinfo {author} {\bibfnamefont {E.}~\bibnamefont {Lescano}}, \bibinfo {author} {\bibfnamefont {G.}~\bibnamefont {Menezes}},\ and\ \bibinfo {author} {\bibfnamefont {J.~A.}\ \bibnamefont {Rodr\'\i{}guez}},\ }\bibfield  {title} {\bibinfo {title} {{Aspects of conformal gravity and double field theory from a double copy map}},\ }\href {https://doi.org/10.1103/PhysRevD.108.126017} {\bibfield  {journal} {\bibinfo  {journal} {Phys. Rev. D}\ }\textbf {\bibinfo {volume} {108}},\ \bibinfo {pages} {126017} (\bibinfo {year} {2023})},\ \Eprint {https://arxiv.org/abs/2307.14538} {arXiv:2307.14538 [hep-th]} \BibitemShut {NoStop}%
\bibitem [{\citenamefont {Lescano}\ and\ \citenamefont {Rodr\'\i{}guez}(2024{\natexlab{a}})}]{Lescano:2024gma}%
  \BibitemOpen
  \bibfield  {author} {\bibinfo {author} {\bibfnamefont {E.}~\bibnamefont {Lescano}}\ and\ \bibinfo {author} {\bibfnamefont {J.~A.}\ \bibnamefont {Rodr\'\i{}guez}},\ }\bibfield  {title} {\bibinfo {title} {{Constructing Conformal Double Field Theory through a Double Copy Map}},\ }\href@noop {} {\  (\bibinfo {year} {2024}{\natexlab{a}})},\ \Eprint {https://arxiv.org/abs/2408.11892} {arXiv:2408.11892 [hep-th]} \BibitemShut {NoStop}%
\bibitem [{\citenamefont {Lescano}\ and\ \citenamefont {Rodr\'\i{}guez}(2024{\natexlab{b}})}]{Lescano:2024lwn}%
  \BibitemOpen
  \bibfield  {author} {\bibinfo {author} {\bibfnamefont {E.}~\bibnamefont {Lescano}}\ and\ \bibinfo {author} {\bibfnamefont {J.~A.}\ \bibnamefont {Rodr\'\i{}guez}},\ }\bibfield  {title} {\bibinfo {title} {{Quadratic Curvature Corrections in Double Field Theory via Double Copy}},\ }\href@noop {} {\  (\bibinfo {year} {2024}{\natexlab{b}})},\ \Eprint {https://arxiv.org/abs/2409.05628} {arXiv:2409.05628 [hep-th]} \BibitemShut {NoStop}%
\bibitem [{\citenamefont {Y\i{}lmaz}(2024)}]{rasim}%
  \BibitemOpen
  \bibfield  {author} {\bibinfo {author} {\bibfnamefont {R.}~\bibnamefont {Y\i{}lmaz}},\ }\bibfield  {title} {\bibinfo {title} {{On the extension of double copy procedure to higher derivative double field theory}},\ }\href@noop {} {\  (\bibinfo {year} {2024})},\ \Eprint {https://arxiv.org/abs/2408.16524} {arXiv:2408.16524 [hep-th]} \BibitemShut {NoStop}%
\bibitem [{\citenamefont {Bonezzi}\ \emph {et~al.}(2022)\citenamefont {Bonezzi}, \citenamefont {Diaz-Jaramillo},\ and\ \citenamefont {Hohm}}]{Bonezzi}%
  \BibitemOpen
  \bibfield  {author} {\bibinfo {author} {\bibfnamefont {R.}~\bibnamefont {Bonezzi}}, \bibinfo {author} {\bibfnamefont {F.}~\bibnamefont {Diaz-Jaramillo}},\ and\ \bibinfo {author} {\bibfnamefont {O.}~\bibnamefont {Hohm}},\ }\bibfield  {title} {\bibinfo {title} {{The gauge structure of double field theory follows from Yang-Mills theory}},\ }\href {https://doi.org/10.1103/PhysRevD.106.026004} {\bibfield  {journal} {\bibinfo  {journal} {Phys. Rev. D}\ }\textbf {\bibinfo {volume} {106}},\ \bibinfo {pages} {026004} (\bibinfo {year} {2022})},\ \Eprint {https://arxiv.org/abs/2203.07397} {arXiv:2203.07397 [hep-th]} \BibitemShut {NoStop}%
\bibitem [{\citenamefont {Bonezzi}\ \emph {et~al.}(2023{\natexlab{a}})\citenamefont {Bonezzi}, \citenamefont {Chiaffrino}, \citenamefont {Diaz-Jaramillo},\ and\ \citenamefont {Hohm}}]{Bonezzi2}%
  \BibitemOpen
  \bibfield  {author} {\bibinfo {author} {\bibfnamefont {R.}~\bibnamefont {Bonezzi}}, \bibinfo {author} {\bibfnamefont {C.}~\bibnamefont {Chiaffrino}}, \bibinfo {author} {\bibfnamefont {F.}~\bibnamefont {Diaz-Jaramillo}},\ and\ \bibinfo {author} {\bibfnamefont {O.}~\bibnamefont {Hohm}},\ }\bibfield  {title} {\bibinfo {title} {{Gauge invariant double copy of Yang-Mills theory: The quartic theory}},\ }\href {https://doi.org/10.1103/PhysRevD.107.126015} {\bibfield  {journal} {\bibinfo  {journal} {Phys. Rev. D}\ }\textbf {\bibinfo {volume} {107}},\ \bibinfo {pages} {126015} (\bibinfo {year} {2023}{\natexlab{a}})},\ \Eprint {https://arxiv.org/abs/2212.04513} {arXiv:2212.04513 [hep-th]} \BibitemShut {NoStop}%
\bibitem [{\citenamefont {Bonezzi}\ \emph {et~al.}(2023{\natexlab{b}})\citenamefont {Bonezzi}, \citenamefont {Chiaffrino}, \citenamefont {Diaz-Jaramillo},\ and\ \citenamefont {Hohm}}]{Bonezzi3}%
  \BibitemOpen
  \bibfield  {author} {\bibinfo {author} {\bibfnamefont {R.}~\bibnamefont {Bonezzi}}, \bibinfo {author} {\bibfnamefont {C.}~\bibnamefont {Chiaffrino}}, \bibinfo {author} {\bibfnamefont {F.}~\bibnamefont {Diaz-Jaramillo}},\ and\ \bibinfo {author} {\bibfnamefont {O.}~\bibnamefont {Hohm}},\ }\bibfield  {title} {\bibinfo {title} {{Weakly constrained double field theory: the quartic theory}}\ }\href@noop {} {} (\bibinfo {year} {2023}{\natexlab{b}}),\ \Eprint {https://arxiv.org/abs/2306.00609} {arXiv:2306.00609 [hep-th]} \BibitemShut {NoStop}%
\bibitem [{\citenamefont {Bonezzi}\ \emph {et~al.}(2024)\citenamefont {Bonezzi}, \citenamefont {Chiaffrino}, \citenamefont {Diaz-Jaramillo},\ and\ \citenamefont {Hohm}}]{Bonezzi4}%
  \BibitemOpen
  \bibfield  {author} {\bibinfo {author} {\bibfnamefont {R.}~\bibnamefont {Bonezzi}}, \bibinfo {author} {\bibfnamefont {C.}~\bibnamefont {Chiaffrino}}, \bibinfo {author} {\bibfnamefont {F.}~\bibnamefont {Diaz-Jaramillo}},\ and\ \bibinfo {author} {\bibfnamefont {O.}~\bibnamefont {Hohm}},\ }\bibfield  {title} {\bibinfo {title} {{Weakly constrained double field theory as the double copy of Yang-Mills theory}},\ }\href {https://doi.org/10.1103/PhysRevD.109.066020} {\bibfield  {journal} {\bibinfo  {journal} {Phys. Rev. D}\ }\textbf {\bibinfo {volume} {109}},\ \bibinfo {pages} {066020} (\bibinfo {year} {2024})},\ \Eprint {https://arxiv.org/abs/2309.03289} {arXiv:2309.03289 [hep-th]} \BibitemShut {NoStop}%
\bibitem [{\citenamefont {Hohm}\ and\ \citenamefont {Kwak}(2011{\natexlab{a}})}]{Hohmhetero}%
  \BibitemOpen
  \bibfield  {author} {\bibinfo {author} {\bibfnamefont {O.}~\bibnamefont {Hohm}}\ and\ \bibinfo {author} {\bibfnamefont {S.~K.}\ \bibnamefont {Kwak}},\ }\bibfield  {title} {\bibinfo {title} {{Double Field Theory Formulation of Heterotic Strings}},\ }\href {https://doi.org/10.1007/JHEP06(2011)096} {\bibfield  {journal} {\bibinfo  {journal} {JHEP}\ }\textbf {\bibinfo {volume} {06}},\ \bibinfo {pages} {096}},\ \Eprint {https://arxiv.org/abs/1103.2136} {arXiv:1103.2136 [hep-th]} \BibitemShut {NoStop}%
\bibitem [{\citenamefont {Hohm}\ and\ \citenamefont {Kwak}(2011{\natexlab{b}})}]{Hohmframe}%
  \BibitemOpen
  \bibfield  {author} {\bibinfo {author} {\bibfnamefont {O.}~\bibnamefont {Hohm}}\ and\ \bibinfo {author} {\bibfnamefont {S.~K.}\ \bibnamefont {Kwak}},\ }\bibfield  {title} {\bibinfo {title} {{Frame-like Geometry of Double Field Theory}},\ }\href {https://doi.org/10.1088/1751-8113/44/8/085404} {\bibfield  {journal} {\bibinfo  {journal} {J. Phys. A}\ }\textbf {\bibinfo {volume} {44}},\ \bibinfo {pages} {085404} (\bibinfo {year} {2011}{\natexlab{b}})},\ \Eprint {https://arxiv.org/abs/1011.4101} {arXiv:1011.4101 [hep-th]} \BibitemShut {NoStop}%
\bibitem [{\citenamefont {Lescano}\ and\ \citenamefont {Rodr\'\i{}guez}(2021)}]{Lescanohetero2}%
  \BibitemOpen
  \bibfield  {author} {\bibinfo {author} {\bibfnamefont {E.}~\bibnamefont {Lescano}}\ and\ \bibinfo {author} {\bibfnamefont {J.~A.}\ \bibnamefont {Rodr\'\i{}guez}},\ }\bibfield  {title} {\bibinfo {title} {{Higher-derivative heterotic Double Field Theory and classical double copy}},\ }\href {https://doi.org/10.1007/JHEP07(2021)072} {\bibfield  {journal} {\bibinfo  {journal} {JHEP}\ }\textbf {\bibinfo {volume} {07}},\ \bibinfo {pages} {072}},\ \Eprint {https://arxiv.org/abs/2101.03376} {arXiv:2101.03376 [hep-th]} \BibitemShut {NoStop}%
\bibitem [{\citenamefont {Cho}\ and\ \citenamefont {Lee}(2019{\natexlab{b}})}]{Chohetero}%
  \BibitemOpen
  \bibfield  {author} {\bibinfo {author} {\bibfnamefont {W.}~\bibnamefont {Cho}}\ and\ \bibinfo {author} {\bibfnamefont {K.}~\bibnamefont {Lee}},\ }\bibfield  {title} {\bibinfo {title} {{Heterotic Kerr-Schild Double Field Theory and Classical Double Copy}},\ }\href {https://doi.org/10.1007/JHEP07(2019)030} {\bibfield  {journal} {\bibinfo  {journal} {JHEP}\ }\textbf {\bibinfo {volume} {07}},\ \bibinfo {pages} {030}},\ \Eprint {https://arxiv.org/abs/1904.11650} {arXiv:1904.11650 [hep-th]} \BibitemShut {NoStop}%
\bibitem [{\citenamefont {Angus}\ \emph {et~al.}(2021)\citenamefont {Angus}, \citenamefont {Cho},\ and\ \citenamefont {Lee}}]{Angus}%
  \BibitemOpen
  \bibfield  {author} {\bibinfo {author} {\bibfnamefont {S.}~\bibnamefont {Angus}}, \bibinfo {author} {\bibfnamefont {K.}~\bibnamefont {Cho}},\ and\ \bibinfo {author} {\bibfnamefont {K.}~\bibnamefont {Lee}},\ }\bibfield  {title} {\bibinfo {title} {{The classical double copy for half-maximal supergravities and T-duality}},\ }\href {https://doi.org/10.1007/JHEP10(2021)211} {\bibfield  {journal} {\bibinfo  {journal} {JHEP}\ }\textbf {\bibinfo {volume} {10}},\ \bibinfo {pages} {211}},\ \Eprint {https://arxiv.org/abs/2105.12857} {arXiv:2105.12857 [hep-th]} \BibitemShut {NoStop}%
\end{thebibliography}%

\end{document}